\documentclass[aps,reprint,superscriptaddress,floatfix,showpacs,pra]{revtex4-2}
\usepackage{graphicx}
\usepackage{amsmath}
\usepackage{amssymb}
\usepackage{amsfonts}
\usepackage{appendix}
\usepackage[unicode=true,breaklinks=true,colorlinks=true]{hyperref}
\hypersetup{citecolor=blue,urlcolor=blue}

\usepackage{newtxtext}
\usepackage{newtxmath}
\usepackage[all]{hypcap}
\usepackage{natbib}
\usepackage{multirow}
\usepackage{booktabs}
\usepackage{bbold}
\usepackage{bm}
\usepackage{bbm}

\newcommand{\ket}[1]{|{#1}\rangle}
\newcommand{\bra}[1]{\langle{#1}|}

\usepackage[dvipsnames]{xcolor}

\usepackage{indentfirst}
\usepackage{derivative}
\usepackage[dvipsnames]{xcolor}\newcommand{\matrixel}[3]{\left\langle{#1}\middle|{#2}\middle|{#3}\right\rangle}

\newcommand{\expval}[1]{\left\langle{#1}\right\rangle}

\begin{document}

\title{Efficiency of optimal control for noisy spin qubits in diamond}
\author{Hendry M. Lim}
\thanks{These two authors contributed equally.}
\affiliation{Department of Physics, Faculty of Mathematics and Natural Sciences, 
Universitas Indonesia, Depok 16424, Indonesia}
\author{Genko T. Genov}
\thanks{These two authors contributed equally.}
\affiliation{Institute for Quantum Optics and Center for Integrated Quantum Science and 
Technology, Universit\"{a}t Ulm, D-89081 Ulm, Germany}
\author{Roberto Sailer}
\affiliation{Institute for Quantum Optics and Center for Integrated Quantum Science and 
Technology, Universit\"{a}t Ulm, D-89081 Ulm, Germany}
\author{Alfaiz Fahrurrachman}
\affiliation{Department of Physics, Faculty of Mathematics and Natural Sciences, 
Universitas Indonesia, Depok 16424, Indonesia}
\author{Muhammad A. Majidi}
\affiliation{Department of Physics, Faculty of Mathematics and Natural Sciences, 
Universitas Indonesia, Depok 16424, Indonesia}
\author{Fedor Jelezko}
\affiliation{Institute for Quantum Optics and Center for Integrated Quantum Science and 
Technology, Universit\"{a}t Ulm, D-89081 Ulm, Germany}
\author{Ressa S. Said}
\email{ressa.said@uni-ulm.de}
\affiliation{Institute for Quantum Optics and Center for Integrated Quantum Science and 
Technology, Universit\"{a}t Ulm, D-89081 Ulm, Germany}
	
\begin{abstract}
Decoherence is a major challenge for quantum technologies. A way to mitigate its negative impact is by employing quantum optimal control. 
The decoherence dynamics varies significantly based on the characteristics of the surrounding environment of qubits, consequently affecting the outcome of the control optimization. In this work, we investigate the dependence of the shape of a spin inversion control pulse on the correlation time of the environment noise. Furthermore, we analyze the effects of constraints and optimization options on the optimization outcome and identify a set of strategies that improve the optimization performance. Finally, we present an experimental realization of the numerically-optimized pulses validating the optimization feasibility. Our work serves as a generic yet essential guide to implementing optimal control in the presence of realistic noise, e.g., in nitrogen-vacancy centers in diamond. 
\end{abstract}

\maketitle
	
\section{Introduction}\label{Section_Introduction}

The potential applications of quantum systems for practical, scalable technologies are hindered by their inevitable interaction with the environment, leading to the loss of \textit{quantumness} known as decoherence. A prominent example is a negatively charged nitrogen-vacancy~$\mathrm{NV}^{-}$ center in diamond, herein referred simply as \textit{the NV center}, known for its long coherence time at room temperature and broad applications for quantum technologies, for instances for quantum sensing~\cite{ Rembold2020, Doherty2013, Prawer2014, BarGill2013, Herbschleb2019,Degen2017, Oshnik2022, Hirose2012, Vetter2022}, quantum computing~\cite{Childress2013, Ladd2010}, quantum networks~\cite{Nemoto2016, Ruf2021}, electric field sensing~\cite{Block2021}, electrochemical sensing~\cite{Dinani2021}, magnetic imaging~\cite{Chipaux2015, Boretti2019}, magnetic resonance spectroscopy~\cite{Abeywardana2014}, masers~\cite{Breeze2018, Wu2021Maser}, relaxometry~\cite{Tetienne2013, Mzyk2022}, gyroscopy~\cite{Brgler2023}, single-photon source~\cite{Lau_2000}, fluorescent biomarkers, thermometry~\cite{Schirhagl2014}, biosensing~\cite{Zhang2021biosensing}, and magnetic biomedical applications~\cite{Kuwahata2020}, supported by accessible synthesis methods such as chemical vapor deposition (CVD)~\cite{Rembold2020, Prawer2014}. Despite its wide applicability, it suffers from decoherence due to interactions with the surrounding spin bath.

The NV center has $S=1$ ground state electron spin \cite{Rembold2020, Aharon2016} and we typically apply a magnetic field, which splits the $S=\pm 1$ ground states, and we can isolate a qubit between, e.g., $S=0$ and $S=-1$ states. Decoherence~\cite{Hanson2008} is typically characterized by a decay time $T_2$, as known in conventional nuclear magnetic resonance~\cite{akcakaya2022magnetic, Blundell2001, XuYao+2019}. It affects the fidelity of pulses that we use for coherent control, e.g., to induce Rabi oscillation or invert the qubit populations for dynamical decoupling (DD) sequences that suppress the effect of decoherence~\cite{Viola1999PRL,Genov2020, Genov2019, Stark2017, wang2012comparison,Suter2016RevModPhys}. One way to improve control fidelity is to perform population inversion operations as fast as possible. However, this is not generally viable, for instance, if the Rabi frequency is limited. Several strategies exist to perform robust population inversion in the presence of pulse errors and decoherence. These include composite pulses \cite{Levitt1986,Wimperis1994,Jones2003PRA,Torosov2011,Genov2013PRL,Genov2014,Hain2020pra,Torosov2022PRL}, rapid adiabatic passage~\cite{Shore1990,Vitanov2001,Tannus97AdiabBook,GarwoodJMR1991,ZlatanovPRA2020,Lauro2011PRA,Mieth2012PRA,Pascual-WinterPRB2012,Genov2020}, continuous driving~\cite{CaiNJP2012,CohenFP2017,Baumgart2016PRL,AharonPRL2013,TimoneyNature2011,Genov2019,Salhov2024prl,Louzon2024arxiv}, stimulated Raman adiabatic passage~\cite{Vitanov1997a,Vitanov2017,Tian2019,Genov2023}, and quantum optimal control methods~\cite{Tian2019,Rembold2020}. 

Our main interest is directed toward optimal control, which uses iterative algorithms to optimize the control pulse based on the control's degrees of freedom. Distinctively, it can be robust against environmental noise using only the control parameters of the pulse itself. There is no need for additional mechanisms such as a dressing field, which may introduce additional constraints to the system and require a deeper analysis. The versatile nature of the optimal control allows it to be highly applicable and adaptable to a wide range of qubit operations, such as quantum sensing~\cite{Rembold2020, Poulsen2022, Gomez2019}, quantum information processing~\cite{Rembold2020, Said2009, Tian2024, Nobauer2015, Tian2020}, and dynamical decoupling~\cite{Poulsen2022, Nobauer2015, Tian2020}. Furthermore, it complements other strategies to provide robust operational performance~\cite{Tian2019, Tian2024}.

The properties of the NV center, e.g. coherence time, vary depending on its environment~\cite{deLange2012, Jamonneau2016, Kim2015, Meinel2023}. It leads us to question how the results from optimal control optimization vary for different environmental noise. Despite the widespread usage, there has so far been no dedicated study that investigates how the parameters of the environmental noise, e.g., correlation time, affect the outcome of the optimization. Specifically, the optimized pulse is found to differ when it is built for different time windows~\cite{Tian2019, Tian2024}, and when the optimization algorithm parameters and goal vary~\cite{Tian2020,Said2009}. It motivates our current work to study and analyze how the parameters of the qubit environment affect the resulting optimized dynamics. While the study has been focused on NV centers, it is also applicable to any other physical system where optimal control is applied to optimize qubit dynamics, e.g., in superconducting qubits and trapped ions.

Our work presents a guideline for optimal control of a qubit under two non-ideal cases: environment-induced decoherence and control amplitude noise. We build a control landscape appropriate for the given problem and use a commonly used optimal control scheme, namely the dressed chopped random basis (dCRAB)~\cite{muller2022} with the Fourier basis option, to optimize the control pulse for spin inversion under different decoherence dynamics represented by Ornstein-Uhlenbeck (OU) noise process \cite{Gillespie1996,Gillespie1996AJP,Uhlenbeck1930,UhlenbeckRMP1945}. We find that the optimized pulse shape varies depending on the correlation time of the decoherence-inducing noise. For a particular correlation time longer than the pulse duration, the pulse is reshaped, so the time correlation of the noise is used to find a robust path for population inversion. In contrast, the optimization produces fast pulses with limited robustness for short correlation times with the optimal pulse approaching a simple rectangular pulse for very short correlation times. In addition, we analyze how the various optimization parameters, both general and scheme-specific, affect the optimization outcomes. As we proceed from fixing the pulse's initial phase to allowing it to take a constant value, the pulse shape significantly changes. Nonetheless, the performance does not change meaningfully. When we modulate both control phase and amplitude, the resulting pulse shape consequently changes yet the performance worsens, implying that using a higher degree of freedom can be detrimental for limited computational resources. We also vary the number of variables to optimize, in which case we observe worse pulse performances due to the more degrees of freedom and the limited computational time. We then vary the number of wiggles in the optimized pulse, producing an improved pulse performance for low correlation times and a mixed performance for the other cases. Finally, we demonstrate our optimization results applied in a real scenario by generating the pulse experimentally for several cases. The experimentally generated pulses fit the numerically optimized counterparts reasonably well, with imperfections prominent in the small and fast-changing parts of the pulses. It reveals the limitations of the actual experimental setup that necessitates further study in the control devices characterization beyond the control optimization~\cite{Singh2023}.

This paper is organized as follows. In Section~\ref{Section_The_System_and_Noise_Process}, we describe the system dynamics starting from the qubit description for the NV center. We then introduce the nonidealities and present the corresponding mathematical descriptions. Also, we visualize how different decoherence dynamics are obtained by varying the parameters from the mathematical descriptions. Section~\ref{Section_The_Optimal_Control} describes the optimal control method. We start by building the control landscape and specifying the system dynamics, control objective, and constraints. Briefly, we review the optimal control scheme of our choice, the dCRAB algorithm. We describe our choice of the control pulse to optimize and conclude the section with discussions on the choice of optimization options. We present the optimization results and the discussions thereof in Section~\ref{Section_Optimization_Results}, which we divide into three parts. The first part discusses the general feature of the pulse as the decoherence dynamics are varied. In the second part, we analyze how the optimization options affect the optimization results. The third part elaborates on the experimental feasibility of the optimized pulses. We discuss challenges and provide outlooks in Section~\ref{Outlook} and conclude our work with~Section~\ref{Conclusion}. 


\section{System and Noise Processes}\label{Section_The_System_and_Noise_Process}

\subsection{Ideal nitrogen-vacancy qubit Hamiltonian}\label{section_ideal_nitrogen_vacancy_qubit_hamiltonian}

When considering a qubit, which can be formed, e.g., between the $m_s=0$ and $m_s=-1$ states of a negatively charged NV center in diamond. The evolution of the qubit is then characterized by a Hamiltonian that is simpler than the full Hamiltonian of the NV center~\cite{Rembold2020}. In the Schr\"{o}dinger picture, the simplified Hamiltonian, which does not include noise terms, takes the form ($\hbar=1$)
\begin{equation}\label{eq:ideal_Hamiltonian}
\begin{split}
    \hat{H}_S(t) = \frac{\omega_0}{2}\hat{\sigma}_z+\Omega_1f(t)\cos{\left[\omega_0t+\phi(t)\right]}\hat{\sigma}_x.
\end{split}
\end{equation}
Here $\omega_0$ is the Larmor frequency corresponding to the energy level spacing between the two levels of the qubit---which we denote $\ket{0}=\begin{pmatrix}0&1\end{pmatrix}^\mathrm{T}$ and $\ket{1}=\begin{pmatrix}1&0\end{pmatrix}^\mathrm{T}$ for the lower and upper state, respectively---given by $\omega_0=2\pi\times2.87\ \mathrm{GHz}\pm \gamma B_z$ where $\gamma=2\pi\times 28\ \mathrm{MHz/mT}$ is the NV center's gyromagnetic ratio and $B_z$ is the strength of the static field $\vec{B}_z$ used to establish the qubit, provided that the qubit is established with the $m_s=0$ and either of the $m_s=\pm 1$ states for the corresponding sign. Qubit manipulation is achieved via Rabi oscillation using a linearly polarized oscillating magnetic pulse $\vec{B}_x=B_xf(t)\cos(\omega t+\phi(t))\vec{x}$ perpendicular to the static field $\vec{B}_z$. The peak Rabi frequency is $\Omega_1=\gamma B_x/2$ and can be modulated by an amplitude modulation function $f(t)$. The pulse frequency $\omega$ is ideally tuned to be resonant with the qubit transition frequency, i.e. $\omega=\omega_0$ so that a total population inversion is possible (see Appendix \ref{AppSec:Rabi_oscillation_vs_detuning}). We also require that $\Omega_1\ll\omega_0$ to avoid strong-driving effects, such as the Bloch-Siegert shift \cite{bloch1940magnetic, Yang2019}. We allow a time-dependent phase $\phi(t)$ to enable phase modulation. In the Bloch sphere representation in the Schr\"{o}dinger picture, Rabi oscillation is represented by a spiraling motion as the Bloch vector of the qubit moves up and down due to the driving field. 

A simpler dynamics can be obtained by moving into the interaction picture with respect to $\hat{U}=\exp\left(i\omega_0\hat{\sigma}_zt/2\right)$, which implies that the interaction picture is a frame rotating at the Larmor frequency of the qubit. Applying the rotating-wave approximation (RWA), we obtain
\begin{equation}\label{eq:ideal_hamiltonian_with_RWA}
\begin{split}
    \hat{H}_I(t) &= \hat{U}\hat{H}_S\hat{U}^\dagger-i\hat{U}\odv{\hat{U}^\dagger}{t} 
    \\
    &= \frac{\Omega_1}{2}f(t)\left[\cos(\phi(t))\hat{\sigma}_x+\sin(\phi(t))\hat{\sigma}_y\right]
\end{split}
\end{equation}
In the interaction picture, the Rabi oscillation is represented as a motion along the latitude of the Bloch sphere without the rotation along the longitude. The RWA approximates the exact dynamics better as the rotation frequency of the rotating frame gets larger~\cite{Burgarth2022oneboundtorulethem}. Since $\omega_0\sim\mathrm{GHz}$, the RWA applies well to the NV qubit.

Throughout this text, we allow $\Omega_1$ to be negative. A rotation with frequency $\Omega_1<0$ about an axis represented by $\phi$ is equivalent to a rotation with frequency $|\Omega_1|$ about the axis represented by $(\phi+\pi)\mathrm{mod}(2\pi)$. In the Bloch sphere representation, a clockwise rotation about the former axis and a counterclockwise rotation about the latter axis are the same operation. 

\subsection{Initialization and state readout}

The NV center is a spin $S=1$ system where the $m_s=0$ state is separated from the $m_s=\pm 1$ states by about $h \times 2.87\ \mathrm{GHz}$, where $h$ is Planck's constant, in the ground state and absence of any external magnetic field. With a static magnetic field present, the ground-state splitting is given by $\hbar[(2\pi\times 2.87\ \mathrm{GHz})\pm\gamma B_z]$ where the second term is the Zeeman splitting with $\gamma\approx 2\pi\times 28\ \mathrm{MHz/mT}$ being the gyromagnetic ratio of the NV center, and $\hbar=h/2\pi$. The ground and excited states are separated by about $h\times 471\ \mathrm{THz}$ ($\lambda\approx637\ \mathrm{nm}$). Additionally, two intermediate singlet states exist between the ground and excited $m_s=0,\pm 1$ states, separated from each other by about $h\times 288\ \mathrm{THz}$ ($\lambda\approx1042\ \mathrm{nm}$)~\cite{Rembold2020, Prawer2014}.

Initialization and readout of NV qubits are done via optical excitations. At thermal equilibrium, the Maxwell-Boltzmann distribution governs the distribution in the three ground states. With a $532\ \mathrm{nm}$ radiation and non-radiative relaxation involving the diamond lattice, the NV center is excited into one of the three excited triplets. This transition is spin-conserving. Within a timescale of the order of $10\ \mathrm{ns}$, the NV center undergoes relaxation back to the ground state. This relaxation happens through two routes. The first route is a spin-conserving $637\ \mathrm{nm}$ radiation, which is preferred by the excited $m_s=0$ state. In the second route, preferred by the other excited states, the system relaxes non-radiatively toward the upper intermediate state, followed by a relaxation toward the lower intermediate state in $\sim1\ \mathrm{ns}$ via a $1042\ \mathrm{nm}$ radiation. This lower state is metastable, with a lifetime in the order of $100\ \mathrm{ns}$ before the system relaxes non-radiatively into the ground $m_s=0$ state. This mechanism allows the NV center to be optically pumped into the ground $m_s=0$ state, allowing initialization of an NV qubit in the $m_s=0$ state. The quantum state of the NV qubit is read out using the same mechanism, based on the property that the $m_s=0$ state prefers to relax radiatively, while the others do not. This implies that the photoluminescence intensity of the $637\ \mathrm{nm}$ radiation depends on the population difference between the spin states~\cite{Rembold2020, Prawer2014, Doherty2013, Acosta2010}, also known as optically detected magnetic resonance~(ODMR). We denote the $m_s=0$ state by $\ket{0}$ and one of the $m_s=\pm 1$ states by $\ket{1}$, making up the NV qubit. 

\subsection{Nonidealities with Ornstein-Uhlenbeck processes}\label{subsec_nonidealities_with_ou}

Our system is subject to two main nonidealities due to different types of noise. The NV qubit is subject to magnetic noise via magnetic dipolar interactions with the surrounding $N$ spins. This noise randomly changes over time due to the random spin flips of the surrounding spins, randomly shifting the Larmor frequency of the qubit, causing dephasing. The resulting dynamics is a coherence decay, as measured using a Ramsey measurement scheme~\cite{Hanson2008, delange2010, deLange2012, Jamonneau2016}. We mathematically model this nonideality using a fully-relaxed Ornstein-Uhlenbeck (OU) process \cite{gillespie1992markov, Gillespie1996MathofBrownian,Fahrurrachman2023}, which can be numerically generated using the following updating formula:
\begin{equation}\label{eq:ou_updating_formula}
\begin{split}
    X(t+\Delta t) = X(t)e^{-(\Delta t)/\tau}+\tilde{n}\sqrt{\frac{c\tau}{2}\left(1-e^{-2(\Delta t)/\tau}\right)}
\end{split}
\end{equation}
where the initial value $X(0)$ is taken from a Gaussian distribution with mean $0$ and variance $c\tau/2$. Equation \eqref{eq:ou_updating_formula} is exact for an arbitrary $\Delta t$. Here $\tilde{n}$ is a number taken from a standard Gaussian distribution, while $c$ and $\tau$ are the OU parameters called the diffusion constant and the relaxation time, respectively. The fully-relaxed OU process has the property that at any given time $t$, the process $X(t)$ is Gaussian distributed with a constant mean of $0$ and variance of $\sigma^2=c\tau/2$. Since $c$ only appears in Eq. \eqref{eq:ou_updating_formula} as $\sigma$, we use $\sigma$ and $\tau$ as the OU parameters. Using $\sigma$ is useful as it contains the information about the variance of $X(t)$ during the OU process at a given time. 

We denote the effect of magnetic noise on the detuning 
 of the NV qubit as $\delta(t)$ and label it \textit{$\delta$ noise}. We choose the OU parameters for the $\delta$ noise to reproduce experimental measurements. To do this, we consider the evolution of the NV qubit under free evolution and a Hahn echo sequence. One physically relevant quantity that can be straightforwardly extracted is the coherence decay time $T_2$, the time it takes for the coherence to fall to $1/e$ its initial value under a specific evolution. The notation used adapts the one used in conventional magnetic resonance due to the similarity between the motion of the Bloch vector of the qubit under the magnetic noise, and the motion of a sample's magnetization under the $T_2$-relaxation~\cite{akcakaya2022magnetic, Blundell2001}. Let $T_2^*$ be the coherence decay time in case of free evolution during a Ramsey measurement and $T_2^\mathrm{HE}$ is the coherence decay time with a Hahn echo sequence. Then, the OU parameters $\sigma_\delta$ and $\tau_\delta$ for $\delta(t)$ are obtained by measuring experimentally $T_2^*$ and $T_2^\mathrm{HE}$ and solving the system of equations \cite{Gillespie1996MathofBrownian,Pascual-WinterPRB2012,Senkalla2024prl,Grimm2024arxiv}
\begin{equation}\label{eq:determine_delta_ou_parameters_1}
\begin{split}
    4\tau_\delta e^{-T_2^\mathrm{HE}/2\tau_\delta}-\tau_\delta \left(e^{-T_2^\mathrm{HE}/\tau_\delta}+e^{-T_2^*/\tau_\delta}+2\right)+T_2^\mathrm{HE}-T_2^*&=0
\end{split}
\end{equation}
for $\tau_\delta$, and then by computing
\begin{equation}\label{eq:determine_delta_ou_parameters_2}
\begin{split}
    \sigma_\delta = \left(\tau_\delta^2 e^{-T_2^*/\tau_\delta}-\tau_\delta^2+\tau_\delta T_2^*\right)^{-1/2}
\end{split}
\end{equation}
(see Appendix \ref{AppSec:Determination_of_OU_parameters_delta} for the derivation). The values for $T_2^*$ and $T_2^\mathrm{HE}$ can be obtained from experiments where the pulses are shorter than the coherence decay times (see, e.g.~\cite{Hanson2008}).

In addition to the $\delta$ noise, we consider an uncertainty in the control pulse amplitude. In an actual experimental setup, the magnetic field amplitude generated by the apparatus may not be exactly what is expected. We simply model this uncertainty, which we denote $\epsilon(t)$ and call it \textit{$\epsilon$ noise}, as a five-percent amplitude error using a fully-relaxed OU process with $\tau_\epsilon=\infty$ and $\sigma_\epsilon=\sqrt{c_\epsilon\tau_\epsilon/2}=0.05$. Here we attribute the $\epsilon$ noise to an inhomogeneous driving field, e.g., in the case of driving an ensemble of NV centers. 

Other than inducing decoherence, the $\delta$ noise hinders the efficacy of any control pulse with a finite width. Hence, the longer it takes for a pulse to operate, the worse its performance is due to such $\delta$ noise. Meanwhile, the $\epsilon$ noise causes a deviation of Rabi frequency in each evolution sample leading to a dephasing accumulated with the rotation phase. Therefore, the effect of the $\epsilon$ noise is more apparent for pulses with high pulse areas (longer durations and/or higher peak Rabi frequencies). Whether the effect of the $\epsilon$ noise dominates over the effect of the $\delta$ noise on the control pulse depends on the pulse specification. We discuss this in more detail in Appendix \ref{AppSec:variation_of_pulse_time_and_phase}.


\begin{table}[!b]

    \centering
    
    \caption{Calculated noise parameters for different $\delta$ noise spectra obtained by fixing the free evolution coherence decay time $T_2^*=0.1\ \mathrm{\mu s}$ of the qubit and by varying the relaxation time $\tau_\delta$ of the $\delta$ noise. The corresponding approximate values of the Hahn echo coherence decay time $T_2^\mathrm{HE}$ and the standard deviation $\sigma_\delta=\sqrt{c_\delta\tau_\delta/2}$ of the $\delta$ noise are obtained from~Eq.~\eqref{eq:determine_delta_ou_parameters_1} and Eq.~\eqref{eq:determine_delta_ou_parameters_2}, under the assumption that the instantaneous Hahn echo pulse does not suffer from the OU noises (see Appendix \ref{AppSec:Determination_of_OU_parameters_delta} for the detail). The corresponding noise spectra are visualized in Fig.~\ref{fig1}.}
    
    \label{Table:noise_spectra}

\begin{ruledtabular}
\begin{tabular}{cccc}
    
$T_2^*\ (\mathrm{\mu s})$ & 
$\tau_\delta\ (\mathrm{\mu s})$ & 
$\sigma_\delta\ (\mathrm{MHz})$ & 
$T_2^\mathrm{HE}\ (\mathrm{\mu s})$ 

\\[2pt]
\colrule

$0.1$  & $100$ & $2\pi\times 2.251$ & $1.8$
\\ 
$0.1$  & $10$  & $2\pi\times 2.255$ & $0.85$ 
\\ 
$0.1$  & $1$  & $2\pi\times 2.288$ &   $0.41$ 
\\ 
$0.1$  & $0.1$  & $2\pi\times 2.624$  & $0.22$  
\\ 
$0.1$  & $0.01$  & $2\pi\times 5.305$  &  $0.2$       
\end{tabular}
\end{ruledtabular}
\end{table}


Subject to both noises,~$\delta(t)$ and~$\epsilon(t)$, the full Hamiltonian becomes
\begin{equation}\label{eq:main_hamiltonian}
\begin{split}
\hat{H}_I(t) &= \frac{\delta(t)}{2}\hat{\sigma}_z
\\
&\quad +\frac{\Omega_1}{2}f(t)\left[1+\epsilon(t)\right]\left[\cos(\phi(t))\hat{\sigma}_x+\sin(\phi(t))\hat{\sigma}_y\right],
\end{split}
\end{equation}
which is typically used to describe the NV qubit dynamics under decoherence and control amplitude error~\cite{Aharon2016, Cai2012, Lei2017, Wu2021, Wang2013, Genov2019, Genov2020}. The evolution of the system, characterized by this stochastic Hamiltonian results in 
a mixed quantum state. The average dynamics of the system can be obtained via a closed system equation of motion without further approximations such as the widely-used Lindblad master equation. The expected observable is given by an average over all the noise realizations. Having a number of samples or trials~$N_\mathrm{sample}$, which can be infinite in principle, our readout quantity corresponds to the sample averaged population inversion, as follows  
\begin{equation}\label{eq:readout_quantity}
\begin{split}
    \overline{\expval{\sigma_z}}&=\frac{1}{N_\mathrm{sample}}\sum_{j=1}^{N_\mathrm{sample}}\expval{\sigma_z}_j\\&=\frac{1}{N_\mathrm{sample}}\sum_{j=1}^{N_\mathrm{sample}}\mathrm{tr}(\rho_j\hat{\sigma}_z),
\end{split}
\end{equation} 
where~$\rho_j(t)$ is obtained by evolving the system with the Hamiltonian given by Eq.~\eqref{eq:main_hamiltonian} for the $j$-th noise sample. The dynamics is simulated via the Liouville-von Neumann equation. We provide a more detailed interpretation of this stochastic Hamiltionan in Appendix~\ref{appsec_interpretation_of_the_stochastic_hamiltonian}.

\begin{figure}[!b]
    \centering
    \includegraphics[width=0.97\linewidth]{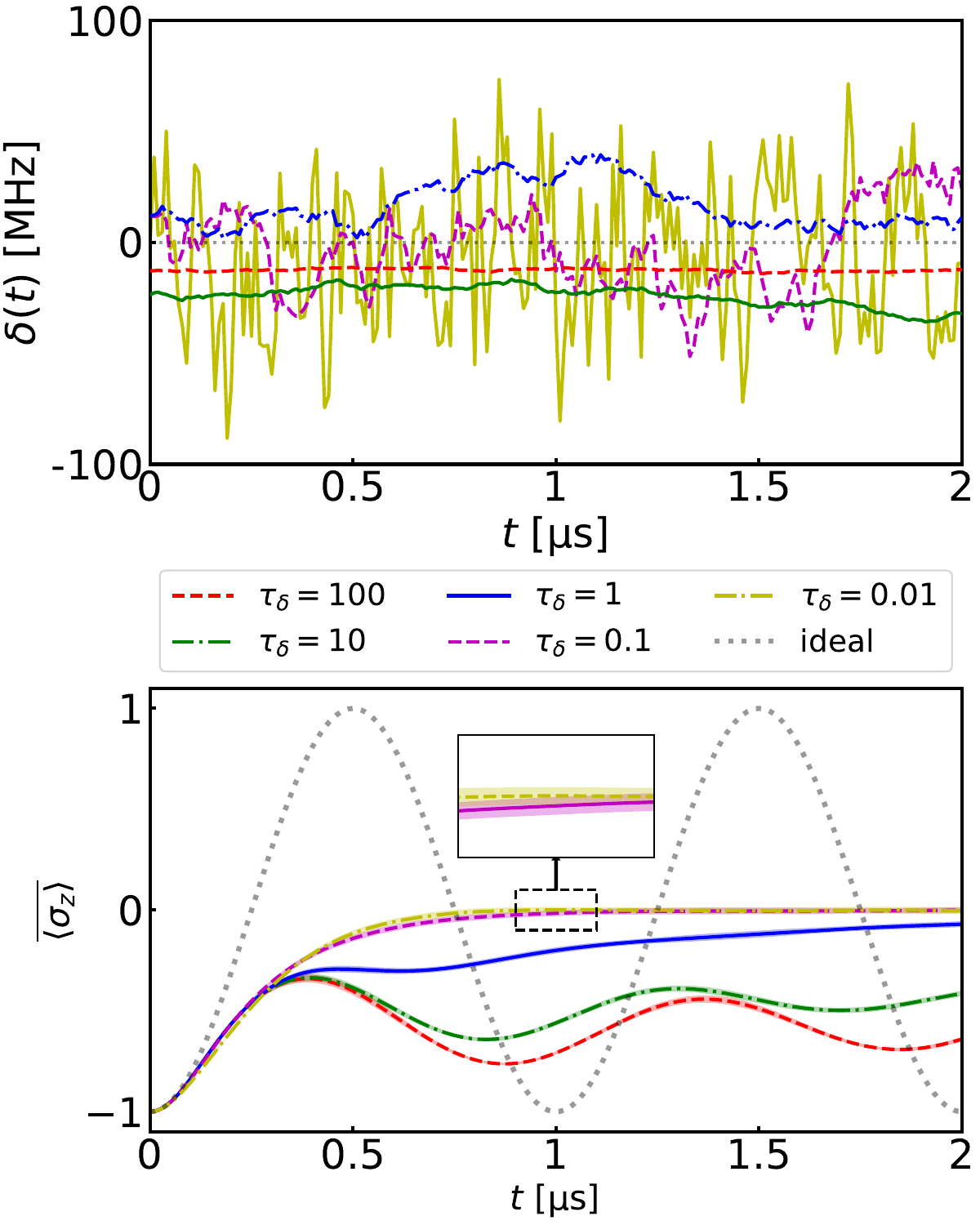}
    \caption{Single random $\delta$ noise samples (top) and the corresponding Rabi oscillation of the sample-averaged population inversion represented by $\overline{\expval{\sigma_z}}$ (bottom). They are obtained by fixing the coherence decay time, $T_2^*=0.1\ \mathrm{\mu s}$, and by varying the correlation time $\tau_\delta$ of the $\delta$ noise. We consider no amplitude noise in this example simulation, i.e.,  $\epsilon=0$. One Rabi period is $2\pi/\Omega_1$, where $\Omega_1=2\pi\,1\ \mathrm{MHz}$. The inset in the bottom figure shows small fluctuations in the average value due to the finite sample size represented by the standard deviation of the fluctuation statistics. It is obtained by repeating the simulation with different noise realizations with $N_\mathrm{sample}=1500, N_\mathrm{rep}=100$. The values for~$\tau_\delta$ are from Table~\ref{Table:noise_spectra}.}
    \label{fig1}
\end{figure}

Numerically, the sample-averaged quantities fluctuate when the calculation is repeated with the different instances of OU noises due to a finite sample size. One needs to repeat the calculations a large number $N_\mathrm{rep}$ of times with the same $N_\mathrm{sample}$ to build the statistics for this ``distribution'' of sample-averaged values. Assuming a normal distribution one should obtain its mean and the corresponding standard deviation. It effectively averages over a larger number~$N_\mathrm{rep}N_\mathrm{sample}$ of noise realizations and provides us with information about the statistical artifact. As we increase $N_\mathrm{sample}$, the average of the distribution converges toward the true average value, while the standard deviation becomes smaller. Throughout this work, we use $N_\mathrm{rep}=100$ and $N_\mathrm{sample}=1500$, which is assumed to be sufficient for the average dynamics as shown by the inset in the bottom part of~Fig.~\ref{fig1}. 

\subsection[Tuning the delta noise]{Tuning the $\delta$ noise}\label{subsec_tuning_the_delta_noise}

The noise parameters for the NV qubit vary due to several factors, for instance, the strength of the interaction with the spin bath~\cite{deLange2012}, the strength of the static field~\cite{Jamonneau2016}, the location of the NV center within the crystal, and the chemical species surrounding the center~\cite{Jamonneau2016, Kim2015, Meinel2023}. Therefore, it is interesting to study the dynamics of the qubit under various noise spectra due to these interactions. Theoretically, one can use Eq.~\eqref{eq:determine_delta_ou_parameters_1} and Eq.~\eqref{eq:determine_delta_ou_parameters_2} to determine the $\delta$ noise spectrum and the respective decoherence dynamics (see also Eq.~\eqref{eq:coherence_under_free_evolution} and Eq.~\eqref{eq:coherence_under_hahn_echo_he}). Here, we are particularly interested in fixing~$T_2^*$ and tuning~$\tau_\delta$, as shown in Table~\ref{Table:noise_spectra}. We obtain $\delta$ noise spectra with progressively larger fluctuation over a given time window (smaller $\tau_\delta$). The spectra also have a more spread-out distribution (i.e.~larger~$\sigma_\delta$), a smaller coherence lifetime recovery (smaller $T_2^\mathrm{HE}$), and significantly worse control pulses. The top part of Fig.~\ref{fig1} shows the single random samples of the different $\delta$ noise spectra and the increase in fluctuations as $\tau_\delta$ decreases. 

We quantify how well a control pulse performs under the given $\delta$ noise by calculating the time it takes for a Rabi oscillation to lose contrast. To show how tuning $\tau_\delta$ affects the control pulses we simulate the Rabi oscillations of $\overline{\expval{\sigma_z}}$ under the different $\delta$ noise spectra without the $\epsilon$ noise and control modulations. The simulation results are presented in the bottom part of Fig.~\ref{fig1}. As the noise correlation time~$\tau_\delta$ decreases, the Rabi oscillation rapidly disappears.

\section{The Optimal Control}\label{Section_The_Optimal_Control}

Here, we describe the quantum optimal control scheme used to optimize the applied pulses by introducing the control landscape description, the optimization algorithm, its options, and the initial pulses to optimize. Throughout the section, the unit of $\tau_\delta$ is in $\mathrm{\mu s}$, and not explicitly written for brevity. 

\subsection{Building the control landscape}

The control amplitude $f(t)$ and phase $\phi(t)$ modulation allow us to freely shape the $x$- and $y$-components in the rotating frame. This enables a control landscape where one can find an optimal pulse that achieves a certain objective by minimizing a corresponding cost function numerically~\cite{cong2014control, Rembold2020}. The landscape, known as the quantum optimal control (QOC) landscape, consists of system dynamics, control objectives, and control space restrictions~\cite{Rembold2020}. Here, the system dynamics is given by Eq.~\eqref{eq:main_hamiltonian}, and the corresponding physical processes are thoroughly discussed in Section~\ref{Section_The_System_and_Noise_Process}. The control objectives are represented by either a cost function~$J$ to be minimized or a figure of merit~$\mathrm{FoM}$ to be maximized. We use the cost function~$J$ throughout the paper. 

Here, we are interested in how the optimized pulse shape changes with respect to the $\delta$ noise spectrum, keeping the dephasing time $T_2^{\ast}$ constant, as mentioned in Sec.~\ref{subsec_tuning_the_delta_noise}. Understanding how the pulse behaves against various noise parameters one obtains an exemplary idea of the optimal control pulse shape for the given noisy spin dynamics. Our optimization objective is focused on but not limited to optimizing an NV spin inversion probability. With the NV spin qubit initially at the state~$\ket{0}$, our goal is to find a pulse that transfers all its population to the state~$\ket{1}$. The appropriate cost function we use here is the closeness between the state at the end of the pulse duration, denoted as~$\rho_j(T)$, and the target state $\rho_\mathrm{target}=\ket{1}\bra{1}$, for a sample~$j$, with $T$~being the pulse duration. Other measures for the cost function can be used for other optimization objectives, e.g.~unitary gate synthesis~\cite{Nielsen2000, Gilchrist2005QuantumDistance, Braunstein1994Bures}. Here, our cost function corresponds to state fidelity~\cite{Nielsen2000}, averaged over the number of samples, $N_\mathrm{sample}$, 
\begin{equation}\label{eq:cost_function}
    J=1-\frac{1}{N_\mathrm{sample}}\sum_{j=1}^{N_\mathrm{sample}}\mathrm{tr}\left(\sqrt{\sqrt{\rho_j(T)}\rho_\mathrm{target}\sqrt{\rho_j(T)}}\right),
\end{equation}
Ideally, $\rho_j(T)=\rho_\mathrm{target}$ for all~$j$, and hence~$J=0$. 

The control space restrictions bound the optimization landscape by limiting the values of the optimized functions $f(t)$ and $\phi(t)$) to avoid unwanted results such as excessive power. As previously mentioned in Sec.~\ref{section_ideal_nitrogen_vacancy_qubit_hamiltonian}, the magnitude of~$\Omega_1$ must also be much smaller than the qubit transition frequency for the rotating-wave approximation to be valid but this condition can in principle be relaxed. For our optimization, we limit the control amplitude $|\Omega_1f(t)|$, not to exceed $10\pi\ \mathrm{MHz}$, at any given time following a typical achievable Rabi frequency for NV centers in diamond ~\cite{Genov2020} with $\Omega_1=2\pi\, 1\,$MHz, therefore $f(t)$ is bounded between $-5$ and $5$.  

\subsection{dCRAB algorithm} 

Various optimal control schemes have been formulated and experimentally implemented. Here, we are particularly interested in the dressed chopped random basis (dCRAB) algorithm as it is one of the methods commonly used in the NV center experiments~\cite{muller2022, Rembold2020}, and was very recently applied for rectifying a quantum gate-set metric experiment with an ensemble of NV centers~\cite{Vetter2024}.

In dCRAB, the optimization is divided into \textit{super-iterations}. In each super-iteration, $f(t)$ and $\phi(t)$ are each expanded into a predetermined---not necessarily the same---number $N_c$ of basis functions. Mathematically, we have
\begin{equation}\label{eq:dCRAB_expansion}
    f^s(t) = a_0^sf^{s-1}(t)+\sum_{k=1}^{N_c}a_k^sf_k^s(\beta_k^s;t)
\end{equation}
where $s=1,2,3,\dots$ and $a_0^1=0$. The expansion is similar for $\phi(t)$. The basis functions $f_k^s(\beta_k^s; t)$ form a complete set in the case where $N_c\rightarrow\infty$. Examples of bases to choose from include the Fourier and Chebyshev bases. Since we only use $N_c$ basis functions, we say that the basis is \textit{chopped}. Each basis function has a randomized parameter $\beta_k^s$, e.g. a random oscillation frequency for a Fourier basis function. Additionally, it is customary to multiply Eq.~\eqref{eq:dCRAB_expansion} by a scaling function $\Lambda(t)$ to determine the general shape of the optimized pulse. We scale our pulse following
\begin{equation}
\Lambda(t)=\tanh\left[\sigma\sin\left(\frac{\pi t}{2T}\right)\right]\tanh\left[-\sigma\sin\left(\frac{\pi(t-T)}{2T}\right)\right],
\end{equation}
where $\sigma=30$ and $T=5\ \mathrm{\mu s}$. This scaling function is equal to unity except near the pulse ends where it quickly decays to zero. We thus set our optimized pulse to always start and end at zero, making it practically convenient.

In the $s$-th super-iteration, the optimization looks for the optimal values of the coefficients $a_0^s, a_k^s$ that minimize $J$ using direct search algorithms such as the Nelder-Mead algorithm. To escape a local minimum, the optimization moves into the next super-iteration with a new set of randomized basis functions. The coefficient $a_0$ lets the optimization move toward the optimized pulse $f^{s-1}(t)$ obtained in the previous super-iteration. By doing this, the optimization is \textit{dressed} with a new search direction.

\subsection{Optimization options}\label{subsec_optimization_options}

One problem we address here is deciding how many variables one needs to optimize. More variables result in a larger control landscape allowing more pulse shape freedom at the expense of more computational cost. For our problem, this concerns whether we optimize either one of or both functions~$f(t)$ and~$\phi(t)$. In particular, for dCRAB it is about how many basis functions we use per super-iteration. It is in principle unclear if more pulse shape parameters guarantee a better optimization result. It may as well be the case that a smaller control space contains already the best-performance pulse shape, given the limited time for the optimization procedure. Therefore, a larger control space may add nothing but a more computational cost. Here, we address such a trade-off between the control space and computational cost by performing multiple rounds of optimizations with varying control landscape sizes. We progressively enlarge the control space following these cases: (1) Optimizing $f(t)$~only while fixing~$\phi(t)=\pi/2$; (2) Optimizing $f(t)$~as a function of time while optimizing $\phi(t)$~as a constant; and (3) Optimizing both $f(t)$ and $\phi(t)$ as functions of time. For the latter two cases, we start with $\phi(t)=\pi/2$ as an initial guess. By increasing the number of parameters this way, we progressively relax the physical constraint imposed on the rotating $x$- and $y$-components of the optimized pulse, respectively: (1) Allowing only the $y$-component; (2) Allowing both the $x$- and $y$-component while constraining their shapes to be identical; and (3) Allowing both components to have different shapes. Also, we vary the number of dCRAB basis functions~$N_c$ per super-iteration for each of the aforementioned cases, varying the numerical constraints of the pulse shapes. For each case, the functions~$f(t)$ and~$\phi(t)$ are expanded with the same~$N_c$.

Other than the control parameter size, there are optimization options in dCRAB that can affect the control landscape. One of these options is the choice of basis functions, which alters the control spaces of some simple optimization problems such that the time it takes for the cost function to converge can significantly differ~\cite {pagano2024rolebasesquantumoptimal}. Here, we use the conventional Fourier basis function to expand our dCRAB for all the optimization cases as it is the typical basis that has been experimentally demonstrated~(see e.g.~works by~\cite{Scheuer_2014, Frank2017}).
For this particular basis, the randomized parameters are the frequencies of the sinusoidal waves. The values are randomly selected from a specified range with a uniform probability distribution. This option is useful since we can directly limit how much the resulting pulse oscillates by bounding the frequency component it may have. Since a lower~$\tau_\delta$ parameter value of the $\delta$ noise leads to a larger fluctuation in the given time window, we hypothesize that the optimized pulse that counteracts the low-$\tau_\delta$ noise is a fast oscillating one. For each above-mentioned case, we vary the upper limit of the random Fourier basis frequencies a pulse can have. We set the lower limit to have a value such that the pulse period is equivalent to~$0.1$ times the resulting sine period. Meanwhile, the upper limit is set to be some variable $\beta_\mathrm{max}$ such that the pulse period is equivalent to $\beta_\mathrm{max}$ times the resulting sine period. Similar to~$N_c$, we set the same $\beta_\mathrm{max}$ for both functions~$f(t)$ and~$\phi(t)$. 

Optimization performances with various options are compared for the same computational resource. The numerical optimization is terminated after~$N_\mathrm{iter}=2000$ total iterations. We set the same convergence condition for each super-iteration. Therefore, the algorithm runs the next super-iteration if~$J$ does no longer improve by~$10^{-3}$ on average over the last 100 iterations, which we assume to be sufficient to indicate that a local minimum has been met.

\subsection{Pulses to optimize}\label{subsection_the_pulse_to_optimize}

The optimization algorithm runs with an initial pulse one wants to optimize. Here, we start the dCRAB with a rectangular pulse having an amplitude of~$\Omega_1=2\pi\,1\, \mathrm{MHz}$, with $f(t)=1$ and $\phi(t)=\pi/2$, applied over a duration~$T=0.5\ \mathrm{\mu s}$. 
The initial control phase is arbitrarily chosen since there is no preferential rotation axis to achieve our goal, while the control pulse width is set such that the pulse performs differently for the values of $\tau_\delta$ given in Table~\ref{Table:noise_spectra}. The pulse will not be affected much by the $\tau_\delta=100, 10$ cases, but significantly affected in the three other cases. As $\tau_\delta=1$ is comparable to the pulse width, while $\tau_\delta=0.1,0.01$ is small compared to the pulse width, we can further divide the cases into $\tau_\delta=1$ and $\tau_\delta=0.1,0.01$ cases. Since the dynamics in the same case group are similar, as shown in Fig.~\ref{fig1}, we hypothesize that the optimized pulse also shows similar properties for cases in the same case group. With our choice of initial pulse, the optimization process may be interpreted as \textit{letting the algorithm reshape the initial rectangular pulse}. It is noteworthy that the $\epsilon$ noise has little effect on this pulse and can be neglected as the total rotation phase is simply $\pi$ (see~Appendix~\ref{AppSec:variation_of_pulse_time_and_phase}). Hence, our results can be explained with only the presence of $\delta$ noise.

The length of time-bins of the pulse is chosen to be as small as possible to allow for more intricate shapes to be accurately designed. Practically, it is the best chosen to match the sampling rate of a waveform generator. The time window of our pulse is divided into 50 time bins, each having a width of~$10\ \mathrm{ns}$. This produces a pulse that has a sampling rate significantly smaller than that of a modern arbitrary waveform generator, which can be as large as $10\ \mathrm{GS/s}$, corresponding to~$0.1\ \mathrm{ns}$ time-bins. However, one needs also to be concerned with a pulse amplifier as it has typically a limited working resolution. Here, we avoid having time bins that are very small to minimize computational resources for our numerical optimizations. Even though it risks a worse resolution, the overall control pulse shape remains approximately the same. 

\section{Optimization Results}\label{Section_Optimization_Results} 

Our optimizations are performed by utilizing a \texttt{Python} numerical package for quantum optimal control, namely Quantum Optimal Control Suite (QuOCS)~\cite{Rossignolo2023}. We run the dCRAB algorithm using the adaptive Nelder-Mead direct search with drift compensation~\cite{Gao2010}. As the package requires additional parameters to be specified, we carry out the following procedures. First, we limit $\phi(t)$ as it is considered a generic function of time. Second, we select large positive and negative numbers as the upper and lower limit, respectively, to consider the periodicity of~$\phi(t)$. Subsequently, we specify an \textit{amplitude variation} for the functions~$f(t)$ and~$\phi(t)$, that control how fast the optimized variables change over the optimization iterations. A very large value may cause the algorithm to frequently skip through a minimum, while a very small value may slow the algorithm in reaching a minimum. Here, we choose the variation for~$f(t)$ and~$\phi(t)$ to be 0.3 times the possible interval width of~$10$ ($\Omega_1f(t)$ is limited at $\pm10\ \mathrm{MHz}$, i.e. $f(t)$ is limited at $\pm 5$), and $2\pi$, respectively. This allows for the dCRAB to traverse the control landscape fast enough with a good convergence towards a global minimum. Finally, as the QuOCS provides an option to either shrink the pulse in each iteration to adhere to the pulse limits or truncate the pulse at the maximum limits, we enable this option to preserve the pulse shape. 

In our optimization runs, the cost function $J$---like any other sample-averaged quantities---fluctuates throughout the optimization iterations. Consequently, the algorithm looks for a good minimum in a fluctuating control landscape. Despite this seemingly unusual feature, optimization in this fluctuating landscape is viable, as shown later by our simulation results. The optimized cost function value is treated with $N_\mathrm{rep}=100$ repetitions with different noise instances. The average value and standard deviation are then reported as $J_\mathrm{opt}$. The standard deviation is found to range from about 0.001 to about 0.008 over all the optimization runs for our choice of parameters. The value is far smaller compared to the corresponding average $J_\mathrm{opt}$, suggesting that the $J$ fluctuation does not significantly affect our optimization results.

As the number of the numerically simulated plots is large, we provide our full optimization results in Appendix~\ref{AppSec:resulting_sigma_z_dynamics_with_optimized_pulses}. Here, Figure~\ref{fig_main} shows the results for several cases, where we show the form of the optimized pulse for each case alongside the corresponding $J_\mathrm{opt}$. Additionally, we present the comparison of $J_\mathrm{opt}$ for all optimization runs in Fig.~\ref{fig_cf_comparison}. The change in~$J$ over the iterations is given in Fig.~\ref{fig_cf_evolution}. The dynamics of~$\overline{\expval{\sigma_z}}$ with the optimized pulses are presented in Fig.~\ref{fig_main_2}. These figures are part of App.~\ref{AppSec:resulting_sigma_z_dynamics_with_optimized_pulses}. For our discussions in this paper, we select only the necessary plots. Furthermore, the values of $J$ for the initial guess rectangular pulse with a duration of $0.5\,\mu$s for the noise spectra with correlation times $\tau_\delta=100,10,1,0.1, 0.001\,\mu$s in Fig. \ref{Table:noise_spectra} are $(0.558\pm 0.008)$, $(0.542\pm 0.008)$, $(0.482\pm 0.007)$,
$(0.392\pm 0.006)$, $(0.378\pm 0.006)$ for $\tau_\delta=100,10,1,0.1,0.01$, respectively.

\begin{figure}[!t]
    \centering
    \includegraphics[width=\linewidth, trim={0 0.1in 0 0}, clip]{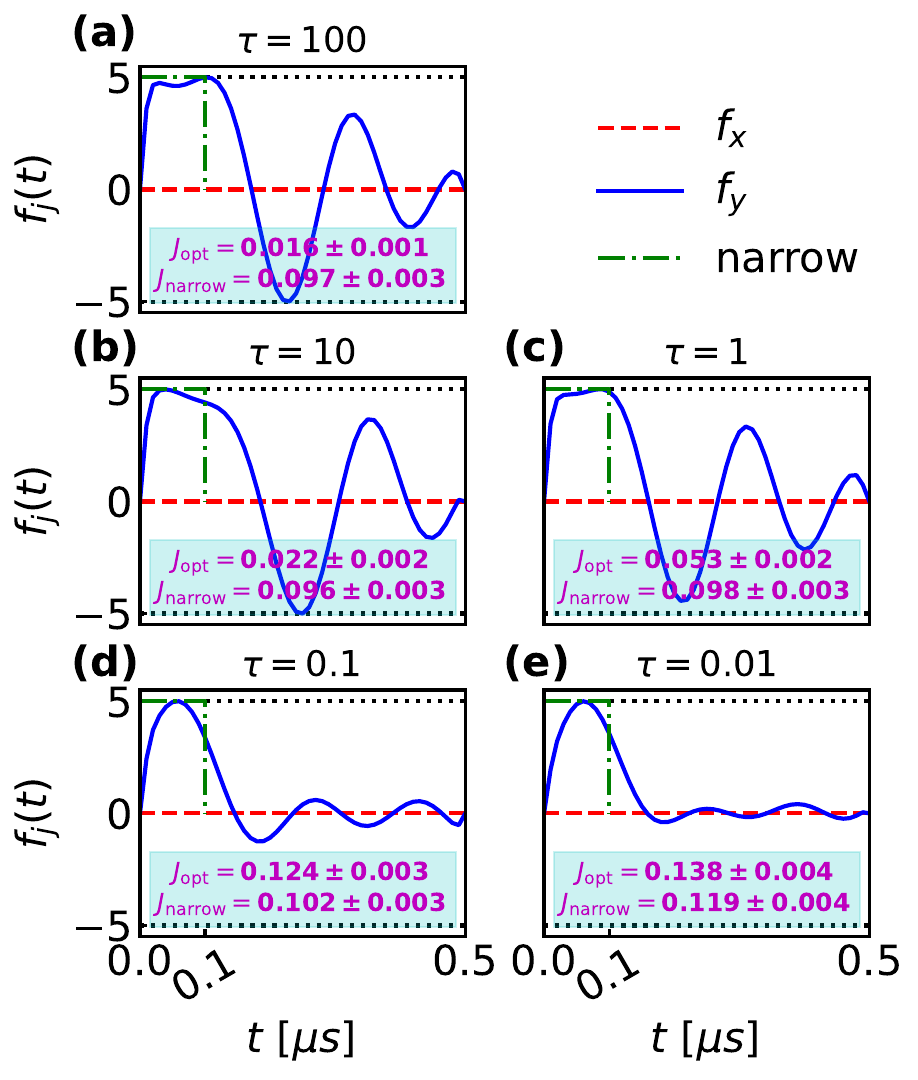}
    \caption{Optimized modulation components $f_x(t)=f(t)\cos[\phi(t)]$ and $f_y(t)=f(t)\sin[\phi(t)]$ vs. the evolution time $t$, for different correlation times of the noise in the detuning $\tau_\delta$. We consider the case where $f(t)$ is optimized with $\phi(t)=\pi/2$, $N_c=5$, and $\beta_\mathrm{max}=3$ for all plots in the figure. On top of each plot, the value of $\tau_\delta$, the optimized cost function $J_\mathrm{opt}$, and the cost function $J_\mathrm{narrow}$ for a short rectangular pulse, which we label \textit{narrow}, are shown. The black dotted lines show the limits in the magnitude of the optimized function. The fixed parameters used here are~$\Omega_1=2\pi\ \mathrm{MHz}, N_\mathrm{iter}=2000, T=0.5\ \mathrm{\mu s}, N_\mathrm{sample} = 1500, N_\mathrm{rep}=100$. These figures are located along the first row of Fig.~\ref{fig_main}~(coded as 1.a.i.)}
    \label{fig2}
\end{figure}

\subsection{General features of the optimized pulses} 

Aside from a few exceptions, our numerical simulations follow a general trend. Contrary to our hypothesis mentioned in Sec.~\ref{subsection_the_pulse_to_optimize}, we observe similar optimized pulse shapes for $\tau_\delta=100, 10, 1$ and $\tau_\delta=0.1,0.01$. It suggests that the resulting pulse shapes generally depend on whether the pulse is applied over a period longer than $\tau_\delta$, instead of how large the total optimization period is compared to $\tau_\delta$. The rectangular pulses are reshaped into non-uniform pulses with unnoticeable defining features, except for a trend that the duration of the main part of the optimized pulses tends to approach the duration of a rectangular $\pi$ pulse at the maximum Rabi frequency  as $\tau_\delta$ decreases. At the same the pulse amplitudes at longer times tend to become smaller. We find this behavior of \textit{squeezing} natural as the most straightforward way to counteract the effect of the $\delta$ noise is to have a control pulse that has the fastest pulse duration. However, the amplitude limit dictates how much the optimized pulses can squeeze themselves. For large~$\tau_\delta$, we observe that the optimized pulses tend to spend some time having negative amplitudes manifesting the direction change of the Rabi oscillation temporarily as shown by Fig.~\ref{fig_main_2}. The cost function improvement indicates that the~\textit{back-and-forth motion} is a valid strategy in counteracting the $\delta$ noise for large $\tau_\delta$. The absence of this feature in low-$\tau_\delta$ cases marks that this feature is unfavorable being outperformed by the squeezing feature.

It is pronounced that all the optimization cases decrease~$J_\mathrm{opt}$, consistently improving the performance of the control pulse. The performance increase depends on $\tau_\delta$. For the sake of clarity, we refer to Fig.~\ref{fig2} to show the optimization results for the case where we optimize $f(t)$, with the fix~$\phi(t)=\pi/2$, $N_c=5$, and $\beta_\mathrm{max}=3$. In most cases, $J_\mathrm{opt}$ increases monotonically as $\tau_\delta$ decreases, i.e., the pulse performance becomes worse with shorter noise correlation time. Comparing this with the original values of~$J$, the reduction in $J$, i.e., its improvement due to optimal control, is less as $\tau_\delta$ becomes smaller. It shows that the faster fluctuations in the $\delta$ noise, which results in worse pulse performances, are non-trivial to counteract. Furthermore, our optimization can obtain lower~$J_\mathrm{opt}$ when $\tau_\delta$ gets larger, as shown by Fig.~\ref{fig_cf_comparison}. 

One may question how the optimized pulses perform compared to the straightforward strategy of using a narrow, high-amplitude pulse. To address this, we calculate~$J$ for a rectangular pulse $f(t)=1$ with $\Omega_1=10\pi\ \mathrm{MHz}$ and $\phi=\pi/2$ applied over $0.1\ \mathrm{\mu s}$, giving a total phase of~$\pi$. Here, we obtain $(0.096\pm 0.003)$, $(0.096\pm 0.003)$, $(0.098\pm 0.003)$, $(0.102\pm 0.004)$, $(0.119\pm 0.004)$ for $\tau_\delta=100, 10, 1, 0.1, 0.01$, respectively. The pulse is depicted in Fig.~\ref{fig2} as the dot-dashed green line. One can clearly notice here that the optimal control pulse improves performance for long correlation times $\tau_\delta$ and the performance becomes slightly worse for the low-$\tau_\delta$ cases. This meets our expectations that the narrow rectangular pulse at the amplitude limit can be considered as the extreme of the squeezing feature. The slightly worse performance of the optimal control pulse can be explained by their specific truncated basis. 

\subsection{Effects of the optimization options}

The optimization options discussed in Sec.~\ref{subsec_optimization_options} do not affect the control behavior in a general way. Nonetheless, those options change some features of the optimized pulse and its performance, which are worthwhile to elaborate on and analyze.

\begin{figure}[!t]
    \centering
    \includegraphics[width=\linewidth]{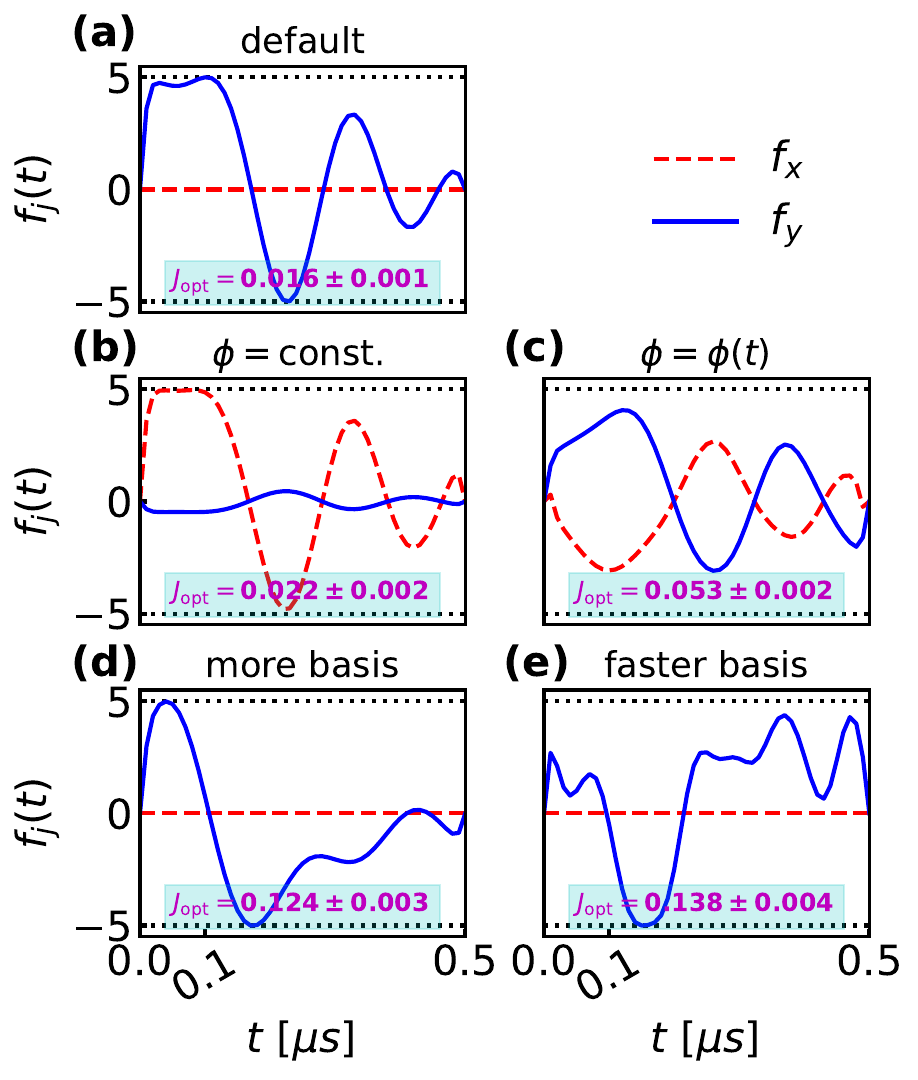}
    \caption{Shape of the optimized pulses vs. the evolution time $t$ for correlation time $\tau_\delta=100\,\mu$s exhibiting how the shape of the optimized pulse and the cost function change as the optimization options are modified. 
    The optimization options parameters$[\phi(t), N_c, \beta_\mathrm{max}]$ take a set of values, as follows: (a) 1.a.i. $[\pi/2, 5, 3]$, (b) 2.a.i. $[\mathrm{const.}, 5, 3]$, (c) 3.a.i $[\phi(t), 5, 3]$, (d) 1.b.i. $[\pi/2, 10, 3]$, (e) 1.a.ii. $[\pi/2, 5, 8]$. The components in (b) have the same phase relation for all $t$. The features of the plots are noticeably similar to Fig.~\ref{fig2}.}
    \label{fig3}
\end{figure}

\subsubsection{Pulse's degrees of freedom}

We observe here a similar change in the optimization outcomes, with the given $\tau_\delta$, $N_c$, and $\beta_\mathrm{max}$, as we relax the constraint on $\phi(t)$. As an example, Figs.~\ref{fig3} (a)--(c) show how the optimized pulse for $\tau_\delta=100, N_c=5,\beta_\mathrm{max}=3$ changes with respect to the $\phi(t)$ constraint. When we allow $\phi(t)$ to take on any constant value, the optimization shifts $\phi(t)$ significantly away from the initial value of $\pi/2$. It results in the nonzero $x$-component as shown in Fig.~\ref{fig3} (b). However, the value of $J_\mathrm{opt}$ does not essentially change. One should expect this result as the $\delta$ noise has no preference in phase. Hence, simply changing $\phi(t)$ will not result in an improved pulse performance and merely add an unnecessary variable to optimize. It is when $\phi(t)$ is allowed to vary in time that we see a change in pulse performance---albeit for the worse. The pulse shape changes significantly as shown by Fig.~\ref{fig3} (c), but the value of $J_\mathrm{opt}$ is higher than when $\phi(t)$ is fixed. By allowing $\phi(t)$ to be time-dependent, we have more freedom that allows us to have a lower $J_\mathrm{opt}$ in principle. However, the resulting control space will be larger. Therefore, it becomes harder for the algorithm to traverse the control landscape and eventually reach a minimum. As shown in Fig.~\ref{fig_cf_evolution}, the transition to the subsequent super-iteration is slow giving a sluggish $J_\mathrm{opt}$ improvement. It is unfavorable to have a time-dependent $\phi(t)$ for a limited computational resource. 

The result is similar for increasing the number of basis functions~$N_c$ per super-iteration. The optimized pulse changes yet the performance remains similar. One will notice increases in $J_\mathrm{opt}$ for the majority of the cases without a meaningful pattern. One example of $J_\mathrm{opt}$ increase is shown in Fig.~\ref{fig3} (a), (d). The larger $N_c$ means more variables to optimize. Hence, the optimization suffers from the same problem as relaxing the $\phi(t)$ constraint. The result is even worse when we use a larger~$N_c$ with a relaxed~$\phi(t)$ constraint, as shown by the last two rows of Fig.~\ref{fig_cf_comparison}. The number of variables to optimize is doubled for the same computational resource.

One may also ask which one is better, to relax the $\phi(t)$ constraint or to increase $N_c$? Our choice of $N_c=10$ for fixed $\phi(t)$ gives only one fewer the number of variables to optimize per super-iteration than allowing $\phi(t)$ to vary with time with $N_c=5$ (see Eq.~\eqref{eq:dCRAB_expansion}). They should serve as good cases for the comparison. Comparing cases (1.b.i) versus (3.a.i), and (1.b.ii) versus (3.a.ii) in Fig.~\ref{fig_cf_comparison}, we see that the advantage of one over the other is inconsistent. One can take, for example, Fig.~\ref{fig3} (c), (d), which indicates that the latter is better, while another case may indicate otherwise. As shown in Fig.~\ref{fig_cf_evolution}, the case of (3.a) optimizations generally goes over fewer super-iterations than the case of (1.b) optimizations. This suggests that allowing $\phi(t)$ to vary with time is better. It takes less basis dressing and therefore a less total number of basis functions to produce a pulse that has a similar performance. The optimization may deal with the same number of variables. However,  relaxing the $\phi(t)$ constraint gives a larger freedom for the physical control pulse, which evidently allows for a trivial decrease in $J_\mathrm{opt}$. It is computationally more expensive to reach convergence in one super-iteration with the additional degrees of freedom.

\subsubsection{Wiggles against wiggles}\label{subsubsection_wiggles}

Here, we discuss the effect of a larger $\beta_\mathrm{max}$. Adding basis functions with higher frequencies, we observe more wiggles in the optimized pulses as shown in Fig.~\ref{fig3} (a), (e). These extra wiggles are arguably more pronounced in the $\tau_\delta=0.1,0.01$ cases despite being small in magnitude compared to the large bump resulting from the squeezing. In the low $\beta_\mathrm{max}$ cases, the optimized pulse is nearly zero over a large portion of the pulse window. In general, the higher-frequency components arise when $\beta_\mathrm{max}$ is increased with small intensities compared to the lower-frequency components. These low-intensity high-frequency components turn out to be beneficial for the optimized pulse performance in some cases. In the $\tau=100, 10, 1$ cases, we see comparable increases and decreases in $J_\mathrm{opt}$, with no particular pattern. For the $\tau=0.1,0.01$ cases, we see consistent decreases in $J_\mathrm{opt}$. Introducing fast-oscillating components here is not useful if~$\tau_\delta$ does not fluctuate much over the pulse window, however, it becomes beneficial if the fluctuations increase. This is in agreement with our previously-mentioned hypothesis (see Sec.~\ref{subsec_optimization_options}). Therefore, increasing $\beta_\mathrm{max}$ provides a way to improve the optimization of the worst-performing low $\tau_\delta$ cases without increasing the number of variables to optimize.

\begin{figure}[!t]
    \centering
    \includegraphics[width=0.75\linewidth]{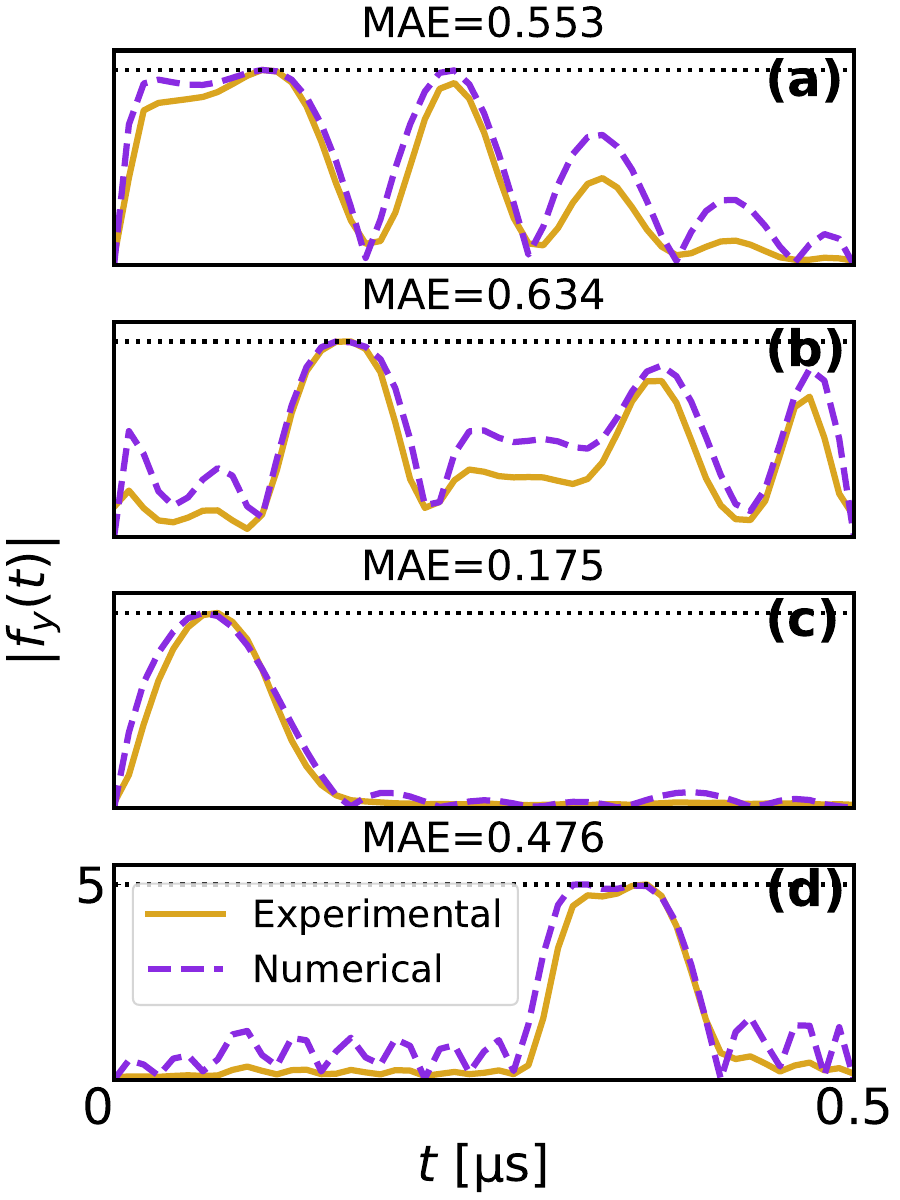}
    \caption{
    Comparison of experimentally and numerically generated optimized pulse shapes $f_y(t)$ ($f_x(t)=0$ for these cases) vs. evolution time $t$ for selected cases with $\phi=\pi/2$ and $N_c=5$: (a) $\tau_\delta=100, \beta_\mathrm{max}=3$ (case 1.a.i.); (b) $\tau_\delta=100, \beta_\mathrm{max}=8$ (case 1.a.ii); (c) $\tau_\delta=0.01, \beta_\mathrm{max}=3$ (case 1.a.i.); (d) $\tau_\delta=0.01, \beta_\mathrm{max}=8$ (case 1.a.ii.). The post-measurement treatment of the experimental pulse is discussed in the main text. On top of each plot on the left column is the mean absolute error~(MAE) of the two pulses.}
    \label{fig4}
\end{figure}

\subsection{Experimental feasibility}

The numerically optimized pulses we have obtained need to be evaluated in a realistic setting to examine their experimental feasibility by sending the pulses through a set of control devices commonly used in experiments with NVs~\cite{Tian2024}. The preset constraint for the maximum pulse amplitude is essential in this context since the devices have an amplitude limit. Also, the amplitude constraint is vital to avoid excessive amplitude, which will damage the micrometer structure of wires on the NV samples. A small variation in the generated pulse shape may significantly affect the pulse performance, so it is necessary to ensure that the pulse shape generation is properly characterized and analyzed. Out of the sixty optimization results shown in Fig.~\ref{fig_main}, we picked out several numerically optimized pulses with different features and sent the pulses to an arbitrary waveform generator (Tektronix AWG70002A with a DAC resolution of 10~bit, $V_{pp}= 500$~mV, and a time resolution of 1~ps), via our homebuilt control suite Qudi~\cite{binder2017qudi,qudi_repo}. The pulses were then generated with a signal amplifier (Amplifier Model No. 60S1G4AM3 AR Germany with a frequency bandwidth of 0.7–4.2~GHz and a gain power of~60~W) and measured by an oscilloscope (LeCroy-WaveAce~234 with a working bandwidth of 300~MHz).

Figure~\ref{fig4} shows the measured pulses compared to the numerical pulses for the selected cases. We take the absolute value of the dCRAB-optimized pulses to match the experimentally generated pulses. Here, a negative amplitude indicates a change of angle by~$\pi$ in the pulse phase~$\phi$~(see Sec.~\ref{section_ideal_nitrogen_vacancy_qubit_hamiltonian}). The experimental pulses are measured in volts over a time-span of $1.2\ \mathrm{\mu s}$ with a step size of $1\ \mathrm{ns}$~(active window of~$0.5\ \mathrm{\mu s}$). The wiggles are due to the mismatch between the sampling rate of the oscilloscope~($500$~MS/s) and the pulse carrier frequency~(2.9~GHz), which corresponds to the frequency~$\omega_0/2\pi$ of the rotating frame. To match the time step of the experimental pulses to the numerical pulses ($0.01\ \mathrm{\mu s}$), and to average over the wiggles, we take the arithmetic mean every ten steps. We shift the experimental pulse vertically so that the minimum value is at zero, then rescale the pulse to match its peak to that of the numerical pulse. We shift the rescaled experimental pulse horizontally to minimize the mean absolute error~(MAE), given by
\begin{equation}\label{eq:exp_opt_diff}
    \mathrm{MAE} = \frac{1}{N_t}\sum_{k=1}^{N_t} \left|f_k^\mathrm{(num)}-f_k^\mathrm{(exp)}\right|,
\end{equation}
where the subscript $k$ indexes the time bin and $N_t=50$ is the number of time bins.

The lowest MAE is attained by the $\tau_\delta=0.01,\beta_\mathrm{max}=3$, which has arguably the simplest shape, as shown by Fig.~\eqref{fig4}~(c). The experimental pulse fit becomes considerably worse for pulses with more complicated features. For the same $\tau_\delta$, the pulses optimized with higher $\beta_\mathrm{max}$ have larger MAE and thus worse experimental fit. Setting the MAE aside, it is obvious that our control devices struggle to generate the small fluctuating components, as prominently shown in Fig.~\ref{fig4}~(d). As discussed in Sec.~\ref{subsubsection_wiggles}, these delicate parts of the pulse may be crucial to oppose the $\delta$ noise with low $\tau_\delta$. The failure to experimentally demonstrate this feature indicates that this strategy may not be viable. Moreover, our devices also struggle to generate pulses consisting of multiple components with spaced-out frequencies as shown by Figs.~\ref{fig4}~(a),~(b).

This result warrants the need to set a control space restriction or to choose the optimization parameters such that the pulses can be well generated within the working capability of the devices. For instance, one can straightforwardly limit the dCRAB randomized basis function parameters reinforcing the constraint. This may lead to a slightly worse, but more experimentally feasible optimization result. Beyond our experimental work here, one may also exploit the measured optimized pulse data and its distortions to obtain the control devices' transfer functions. Such transfer functions can simply be incorporated into the optimization algorithm hence improving the optimization significantly. An effective procedure that is suitable for nonlinear transfer functions of arbitrary control pulse lengths and magnitudes does exist and can be utilized for this purpose~\cite{Singh2023}.

\section{Discussion and Outlook}\label{Outlook}

Here, we briefly note potential future explorations and challenges arising from our current work. 
As shown above, in our example, which is typical for single NV centers in diamond, the $\delta$ noise dominates the spin dynamics, so the pulses are optimized to counteract it. It is interesting to study the dynamics of a particular system that suffers predominantly from the $\epsilon$ noise and to analyze how the optimization can be performed in such a scenario. Large $\epsilon$ noise is typical for ensembles of NV centers due to amplitude inhomogeneity although its spectrum is typically determined by the arbitrary waveform generator and the amplifier used to generate the pulses. Is is also possible to consider another way of optimizing the functions~$f(t)$, and~$\phi(t)$, for example by expanding them as~$f_x(t)=f(t)\cos[\phi(t)]$ and~$f_y(t)=f(t)\sin[\phi(t)]$ in the dCRAB basis and optimizing their expansion coefficients. This alternative may result in a better pulse performance as the optimized pulse features are inherently present in the pulse components. As for the optimization parameters, using only two variations for a given parameter is insufficient to fully establish the dependency. A broader range of dependencies of the dCRAB optimization on the number of bases per super-iteration and the maximum randomized basis parameters is expected to provide a better insight into the best parameter values to use. Increasing the function evaluation limit may also allow for optimizations with a larger number of variables to potentially reach a lower cost function value, provided that we can afford a higher computational cost. To obtain a faster convergence given a limit of computational resources, utilizing other choices of bases may be beneficial~\cite{pagano2024rolebasesquantumoptimal}. Fine-tuning the amplitude variation as one of the features of the optimization package~QuOCS may also yield a faster convergence. The choice of the initial guess pulse can also be critical. For instance, one may start with zero amplitude (no pulse) and let the algorithm build the optimized pulse ground-up. As another example, we may do a \textit{two-step optimization}, where the initial value is taken from the previous optimization with different parameters. The optimization outcome may also differ depending on the control objective. There are various possible expressions for $J$ we may choose, depending on the goal. For example, one may use the gate fidelity when optimizing pulses for dynamical decoupling purposes.

Lastly, the fluctuating $J$ should be considered if we aim for a value lower than the fluctuations, e.g. $J<10^{-3}$, as the fluctuation is likely to \textit{confuse} the algorithm. 
One straightforward solution we propose is to fix the OU noise realization used throughout the optimization, thus holding the control landscape still as it usually is. Since this means that the optimized pulse is tailored for one specific realization, we need to choose the realization such that the corresponding dynamics are as close as possible to the true dynamics. This can be done by repeating the initial $J$ calculation $N_\mathrm{rep}$ times to build the $J$ fluctuation statistics, then choosing the noise realization that produces $J$ that is close to the average of the statistics. This gives us the \textit{close-to-true} control landscape. Presumably, the optimized pulse with the same noise realization will similarly produce $J$ that is close to the average of the $J_\mathrm{opt}$ fluctuation statistics, thus representative of the actual dynamics.

\section{Conclusion}\label{Conclusion}

We have performed control optimization of qubit operations with environment-induced decoherence and control amplitude noise using the dCRAB method, demonstrating the experimental feasibility of such optimized operations. The optimizations improve the fidelity of spin qubit inversion, and the optimized control pulses outperform the standard non-optimized rectangular control pulses. We have shown that the choice of optimization parameters and control constraints are important for the outcome of the optimization procedure and the experimental realization. Our work serves as a guide and demonstrates what the expected forms and characteristics of the optimized pulses are, given realistic experimental constraints in a noisy spin environment.

\vspace{4.00mm}
{\it Acknowledgments.} H.M.L thanks Mahameru BRIN HPC for their facilities. F.J and R.S.S acknowledge DFG, BMBF (CoGeQ, SPINNING, QRX), QC-4-BW, Center for Integrated Quantum Science and Technology, QTBW, ERC Synergy Grant HyperQ, and EU Projects (Spinus, C-QuENS) for their supports. R.S.S thanks Jingfu Zhang, Isabell Jauch, Rajan Paul, and Jiazhao Tian for their discussions.

{\it Author contributions.} G.T.G, F.J and R.S.S conceptualized the work. H.M.L and A.F wrote codes for the dynamical simulations. H.M.L developed the simulation and optimization libraries and performed the numerical work. G.T.G and H.M.L worked on the theoretical analyses. R.S and R.S.S conceived the experiment. R.S performed the measurements for the experimental pulses. H.M.L, G.T.G, R.S, F.J, and R.S.S analyzed the data and drafted the manuscript. M.A.M and F.J acquired the project funding. M.A.M, F.J, G.T.G, and R.S.S supervised the project. All authors read and contributed to the manuscript.

{\it Data Availability Statement.} The datasets in this work are available from the corresponding author upon reasonable request. Processed data
for the plots is publicly available~\cite{plotdata}.

\appendix
	
\section{Rabi oscillation and detuning}\label{AppSec:Rabi_oscillation_vs_detuning}
The Rabi oscillation is ubiquitous in standard textbooks discussing two-level systems (e.g. see~\cite{Griffiths2018, gerry_knight_2004}. Particularly, with Eq.~\eqref{eq:ideal_hamiltonian_with_RWA}, we have
\begin{equation}\label{eq:rabi_osc_sigma_z}
    \left\langle \sigma_z\right\rangle = \frac{\Omega_1^2-\Delta^2}{\Omega_1^2+\Delta^2}\sin^2\left(\frac{\sqrt{\Omega_1^2+\Delta^2}}{2}t\right)-\cos^2\left(\frac{\sqrt{\Omega_1^2+\Delta^2}}{2}t\right),
\end{equation}
when the qubit starts in the $\ket{0}=\begin{pmatrix}0&1\end{pmatrix}^\mathrm{T}$ state, corresponding to $\expval{\sigma_z}_{t=0}=-1$. It is evident that~$\expval{\sigma_z}=1$ is not possible for any $t$ unless the detuning~$\Delta=\omega-\omega_0$ is zero. 

\section{Determination of the OU parameters for the $\delta$ noise}\label{AppSec:Determination_of_OU_parameters_delta}

Firstly, we note that the derivations in this section have previously been done (see Ref.~\cite{pascual-winter2012}, which is our inspiration for our derivation here). We consider the qubit coherence~$\rho_{12}$ under the free evolution. Using Eq.~\eqref{eq:main_hamiltonian} with $\Omega_1=0$, and von Neumann equation $\dot{\rho}=-i[\hat{H},\rho]$, we obtain
\begin{equation*}
\begin{split}
    \dot{\rho}_{12} = -i\delta\rho_{12},
\end{split}
\end{equation*}
whose solution is 
\begin{equation*}
    \rho_{12}(t)=\rho_{12}(0)\exp\left[-i\int_0^t\delta(t')\mathrm{d}t'\right].
\end{equation*}
Taking the average over experimental runs, and noting that $\rho_{12}(0)$ is the same for all runs, we have
\begin{equation*}
    \overline{\rho_{12}}(t)=\rho_{12}(0)\overline{\exp\left[-i\int_0^t\delta(t')\mathrm{d}t'\right]}.
\end{equation*}
For a Gaussian function $G(t)$, and an arbitrary function $f(t)$, 
\begin{equation*}
\overline{\exp\left[i\int_{t_0}^tf(t')G(t')\mathrm{d}t'\right]} = \exp\left[-\frac{1}{2}\overline{\left(\int_{t_0}^t f(t')G(t')\mathrm{d}t'\right)^2}\right].
\end{equation*}
We identify $t_0=0, f(t')=-1, G(t')=\delta(t')$ to get
\begin{equation*}
    \overline{\rho_{12}}(t)=\rho_{12}(0)e^{-\gamma(t)},
\end{equation*}
where
\begin{equation*}
\begin{split}
    \gamma(t)&= \frac{1}{2}\overline{\left(\int_0^t(-1)\delta(t')\mathrm{d}t'\right)^2}
    \\ &= \frac{1}{2}\int_0^t \mathrm{d}t' \int_0^t \mathrm{d}t''\mathrm{cov}\left\{\delta(t'),\delta(t'')\right\}.
\end{split}
\end{equation*}
The covariance $\mathrm{cov}\left\{X(t'),X(t'')\right\}$ of a fully-relaxed OU process $X$ is $\sigma^2e^{-|t'-t''|/\tau}$, where $\sigma^2=c\tau/2$. Furthermore, since $\mathrm{cov}\{X,Y\}=\mathrm{cov}\{Y,X\}$ we use the property that
\begin{equation*}
    \int_0^t\mathrm{d}t'\int_0^t\mathrm{d}t'' f(t',t'')=2\int_0^t\mathrm{d}t'\int_0^{t'}\mathrm{d}t'' f(t',t''),
\end{equation*}
given that $f(t',t'')=f(t'',t')$. Using these, and working out the integral, we obtain
\begin{equation*}
    \gamma(t)=\sigma_\delta^2\tau_\delta^2\left(\frac{t}{\tau_\delta}+e^{-t/\tau_\delta}-1\right),
\end{equation*}
and thus
\begin{equation}\label{eq:coherence_under_free_evolution}
    \overline{\rho_{12}}(t)=\rho_{12}(0)\exp\left[-\sigma_\delta^2\tau_\delta^2\left(\frac{t}{\tau_\delta}+e^{-t/\tau_\delta}-1\right)\right].
\end{equation}

For the evolution under a Hahn echo sequence, we note that the coherence is conjugated each time a $\pi$-pulse is applied, which rotates the Bloch vector by $\pi$ over some axis in the $xy$-plane of the Bloch representation. Let~$T$ be the readout time, so the $\pi$-pulse is applied at $T/2$. Let the pulse be \textit{instantaneously} applied with some arbitrary phase shift $\phi=\alpha$ (in particular, $\phi=0$ or $\phi=\pi/2$ is actually used for a Hahn echo sequence). Then, 
\begin{equation*}
    \rho_{21}^\mathrm{(after\ pulse)} = e^{i(\pi-2\alpha)}\rho_{12}^\mathrm{(before\ pulse)}
\end{equation*}
That is, the coherence $\rho_{12}$---evolving like how a ``$12$'' element evolves before the $\pi$-pulse---``occupies'' the ``$21$'' element after the $\pi$-pulse and evolves like the ``$21$'' element, i.e. under the differential equation $\dot{\rho}_{21}=i\delta\rho_{21}$. Thus, the full equation of motion under the Hahn echo sequence reads
\begin{equation*}
    \dot{\rho}_{12}^\mathrm{HE} = \begin{cases}-i\delta\rho_{12}^\mathrm{HE}, & 0\leq t\leq T/2 \\ i\delta\rho_{12}^\mathrm{HE},  & T/2\leq t\leq T\end{cases},
\end{equation*}
where we have omitted $e^{i(\pi-2\alpha)}$ from both sides of the $T/2<t<T$ case, showing that the choice of $\phi$ is irrelevant for the coherence. 

The solution to this differential equation at readout is
\begin{equation*}
    \rho_{12}^\mathrm{HE}(T) = \rho_{12}(0)\exp\left[-i\int_0^{T/2}\delta(t')\mathrm{d}t'+i\int_{T/2}^{T}\delta(t'')\mathrm{d}t''\right]
\end{equation*}
The rest of the derivation proceeds similarly to the free evolution case. We obtain
\begin{equation}\label{eq:coherence_under_hahn_echo_he}
    \overline{\rho_{12}^\mathrm{HE}}(T)=\rho_{12}(0)\exp\left[-\sigma_\delta^2\tau_\delta^2\left(\frac{T}{\tau_\delta}+4e^{-T/2\tau_\delta}-e^{-T/\tau_\delta}-3\right)\right]
\end{equation}

We then set the exponents in Eqs.~\eqref{eq:coherence_under_free_evolution} and~\eqref{eq:coherence_under_hahn_echo_he} to be equal to one when the coherence lifetimes $T_2^*$ and $T_2^\mathrm{HE}$, respectively, are reached. Simultaneously solving these equations by substituting for $\sigma_\delta^2$ yields Eq.~\eqref{eq:determine_delta_ou_parameters_1} which can be solved to obtain $\tau_\delta$. The value of $\tau_\delta$ can then be substituted to either of Eqs.~\eqref{eq:coherence_under_free_evolution} and~\eqref{eq:coherence_under_hahn_echo_he} to obtain $\sigma_\delta^2$, and hence $c_\delta$. Equation~\eqref{eq:determine_delta_ou_parameters_2} uses the former. 

The requirement that $T_2^*$ and $T_2^\mathrm{HE}$ are obtained with short pulses, discussed in the main text, is a consequence of the assumption of ideal pulses in this derivation.

\section{Interpretation of the stochastic Hamiltonian}\label{appsec_interpretation_of_the_stochastic_hamiltonian}

Usually, the observed dynamics of the system are represented by the state vector $\ket{\psi}$ or the density matrix $\rho$. The expectation value of an observable $\mathcal{O}$ is obtained by calculating $\matrixel{\psi}{\hat{\mathcal{O}}}{\psi}$ or $\mathrm{tr}(\rho\hat{\mathcal{O}})$, which we can interpret as the average outcome when we repeat the measurements over an ensemble of identically prepared systems. However, with the presence of the OU noises, each sample of the system evolves differently depending on the exact form of the noises. For the $j$th evolution sample, the system evolves under the Hamiltonian $\hat{H}_j$ with the $j$th sample of the OU noises. We can expect to measure $\expval{\mathcal{O}}_j = \mathrm{tr}(\rho_j\hat{\mathcal{O}})=\matrixel{\psi_j}{\hat{\mathcal{O}}}{\psi_j}$, where $\rho_j=\ket{\psi_j}\bra{\psi_j}$ is the density matrix evolving with $\hat{H}_j$ under the Schr\"{o}dinger or the Liouville-von Neumann equation. Over the possible noise realizations, we expect to observe the average over these dynamics as given by the sample-averaged expected value $\overline{\expval{\mathcal{O}}}=\sum_{j=1}^{N_\mathrm{sample}} \expval{\mathcal{O}}_j/N_\mathrm{sample}$, where $N_\mathrm{sample}$ is the (large) number of samples. We note that
\begin{equation}
\begin{split}
\overline{\expval{\mathcal{O}}} = \frac{1}{N_\mathrm{sample}}\sum_{j=1}^{N_\mathrm{sample}}\expval{\mathcal{O}}_j 
&= \mathrm{tr}\left(\left(\frac{1}{N_\mathrm{sample}}\sum_{j=1}^{N_\mathrm{sample}}\rho_j\right)\hat{\mathcal{O}}\right)
\\&= \mathrm{tr}(\overline{\rho}\hat{\mathcal{O}})
\end{split}
\end{equation}
where we have defined the \emph{sample-averaged density matrix} $\overline{\rho}$ which describes the dynamics expected over all the noise realizations. 

Evidently, $\overline{\rho}$ and our discussion so far constitute the description of the dynamics of a mixed state~\cite{Schlosshauer2007, miller2008quantum}, making $\overline{\rho}$ the mixed-state density matrix. We do not know which OU noise realization the qubit evolves under for the given evolution run. However, we do know that there is a $1/N_\mathrm{sample}$ chance that the noises are one set among the possible realizations and that our qubit evolves into the corresponding final (pure) state. We may also view an instance of Hamiltonian the system evolves in as an element of a \emph{Hamiltonian ensemble}~\cite{Chen2021ensemble}---the final state $\hat{\rho}$ is the average result of the system's evolution under the Hamiltonian ensemble. 

\begin{figure}[!t]
    \centering
    \includegraphics[width=\linewidth]{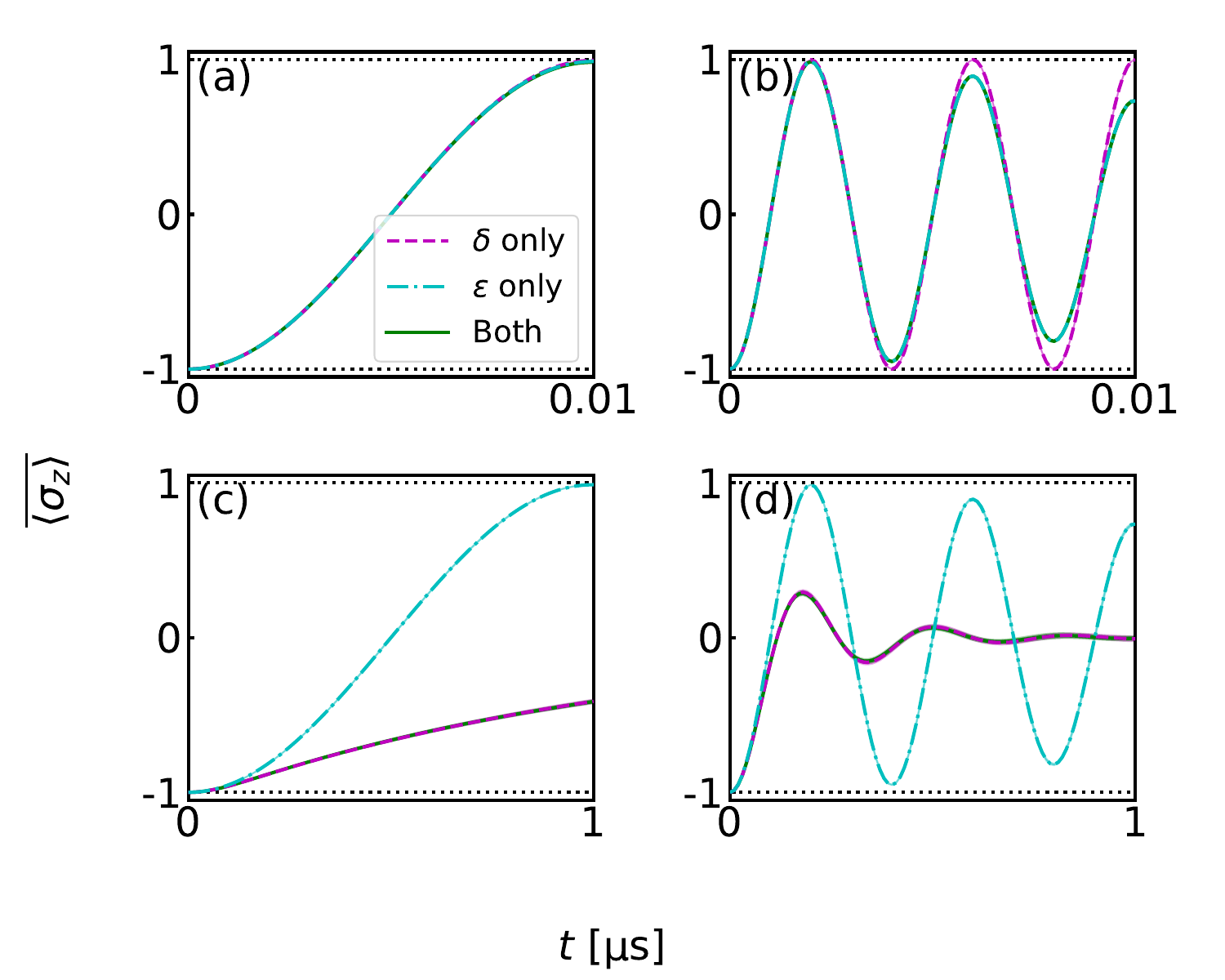}
    \caption{Evolution of $\overline{\expval{\sigma_z}}$ in the presence of different noise sources--- $\delta$ noise only, $\epsilon$ noise only, and both---for rectangular control pulses under different conditions: (a) short duration, small rotation phase; (b) short duration, large rotation phase; (c) long duration, small rotation phase; (d) long duration, long rotation phase. The value $\phi(t)=\pi/2$ is arbitrarily chosen and is the same for all simulations. The OU parameters of the $\delta$ noise, $\tau_\delta=0.1\ \mathrm{\mu s}$ and  $T_2^*=0.1\ \mathrm{\mu s}$, are arbitrarily chosen such that the chosen pulse lengths and rotation phases exhibit different effects. Simulated with $N_\mathrm{sample}=1500,N_\mathrm{rep}=100$.}
    \label{fig_appc}
\end{figure}

\section{The $\delta$ noise versus the $\epsilon$ noise}\label{AppSec:variation_of_pulse_time_and_phase}

As we illustrate in Fig.~\ref{fig1}, the $\delta$ noise also affects the working of a control pulse, apart from causing decoherence. One question to address is whether the effect of the $\delta$ noise dominates over the $\epsilon$ noise for the given pulse. We study how the $\delta$ and $\epsilon$ noise separately affect a control pulse. A control pulse is characterized by two of the following properties: the rotation phase, pulse duration, and pulse frequency, from which the third follows. Here we consider the evolution of $\overline{\expval{\sigma_z}}$ under pulses of different duration and total rotation phases. The simulation is done with only one OU noise present, then with both present. The parameters of the $\delta$ noise and the initial phase $\phi$ of the control pulse are arbitrarily chosen. The values of the pulse properties are chosen accordingly to exhibit changes in behaviors of interest. 

The simulation results are shown in Fig.~\ref{fig_appc}. Figure~\ref{fig_appc} (a) shows that when the pulse is short and does only a small rotation, the dynamics are not affected to any noticeable degree, and the qubit evolves very close to the ideal case where $\overline{\expval{\sigma_z(t)}}=-\cos(\Omega_1 t)$ (see Eq. \eqref{eq:rabi_osc_sigma_z}). As the rotation phase is increased, as shown in Fig.~\ref{fig_appc} (b), we see a deviation in the ``$\epsilon$ only'' case and the ``both'' case, while the ``$\delta$ only'' case stays close to the ideal case. This shows that the effect of the $\epsilon$ noise is dominant for large rotation phases. Moving on to Fig.~\ref{fig_appc} (c), we see that when the rotation phase is small but the pulse duration is long, the $\delta$ noise is dominant. Finally, Fig.~\ref{fig_appc} (d) shows the combined effect of both OU noises. In this case, we see a small deviation of the ``both'' case from the ``$\delta$ only'' case, showing that the $\delta$ noise dominates over the dynamics in case (d). However, we cannot generally tell the dominance of one noise over the dynamics, as the dominance shown in Fig.~\ref{fig_appc} (d) may be caused by the fact that the pulse duration is ten times that of $\tau_\delta$. This indeed makes an interesting problem in and of itself.

\section{Full view of the optimization results}\label{AppSec:resulting_sigma_z_dynamics_with_optimized_pulses}

\begin{figure*}\label{fig_main}
    \centering
    \includegraphics[height=0.8\textheight]{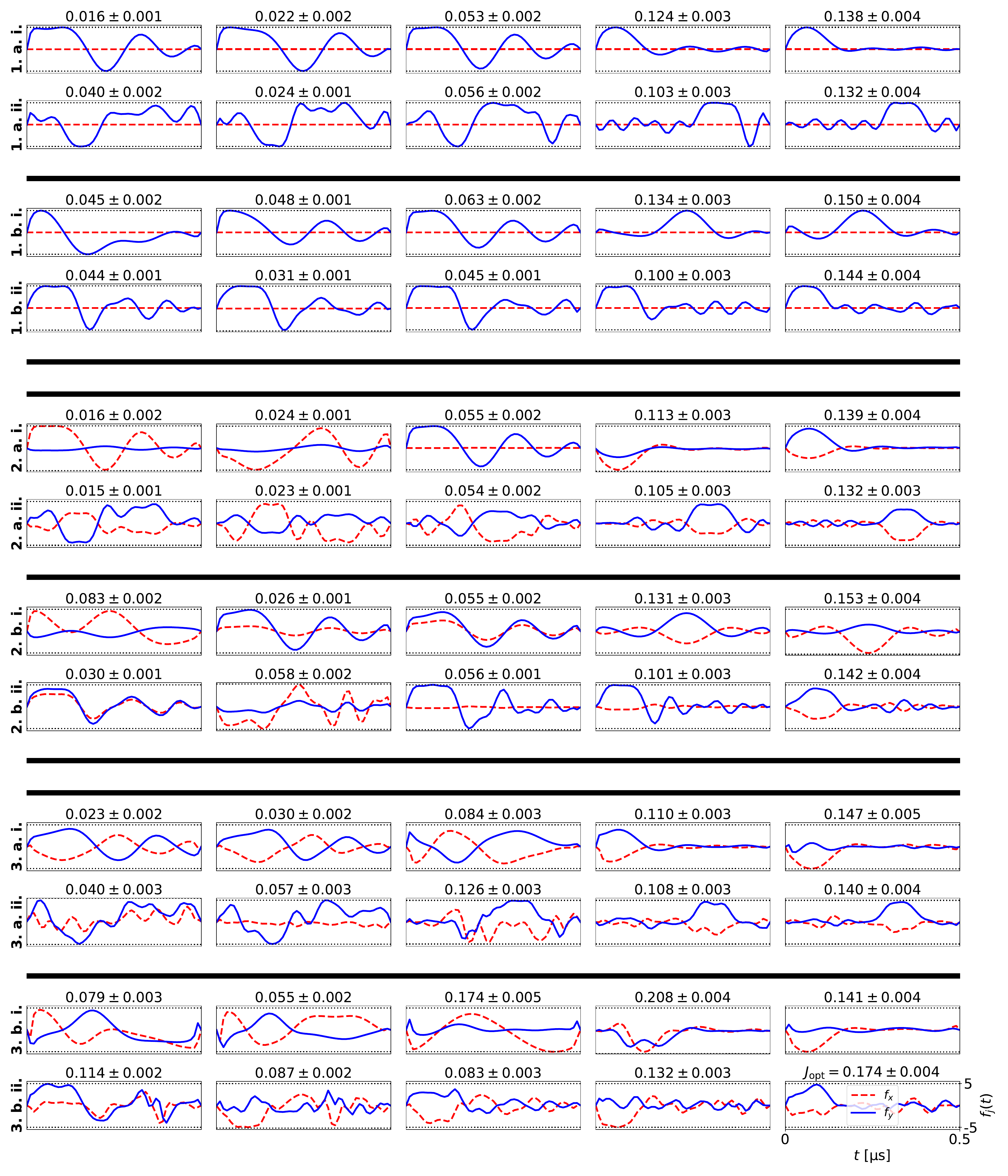}
    \caption{dCRAB-optimized modulation components $f_x(t)=f(t)\cos[\phi(t)]$ and $f_y(t)=f(t)\sin[\phi(t)]$ for different cases specified in Section \ref{Section_The_Optimal_Control}. Each row corresponds to one case denoted by the numberings to the left: \textbf{1.} optimizing $f(t)$ only while fixing $\phi(t)=\pi/2$; \textbf{2.} optimizing $f(t)$ as a function of time while optimizing $\phi(t)$ as a constant; \textbf{3.} optimizing both $f(t)$ and $\phi(t)$ as functions of time; \textbf{a.} five basis functions per superiteration, $N_c=5$; \textbf{b.} $N_c=10$; \textbf{i.} minimum period of the randomized sine wave is a third of the pulse window, $\beta_\mathrm{max}=3$; \textbf{ii.} $\beta_\mathrm{max}=8$. Meanwhile, each column corresponds to each value of $\tau_\delta$; from left to right: $100, 10, 1, 0.1, 0.01$. The best pulse shapes after $N_\mathrm{iter}=2000$ total iterations are shown along with the corresponding cost function value $J_\mathrm{opt}$, rounded to three decimal places. The black dotted lines show the resultant limit corresponding to the amplitude limit of $\pm 10\pi\ \mathrm{Mhz}$. The pulse duration is fixed at $T=0.5\ \mathrm{\mu s}$. The value of $J_\mathrm{opt}$ shown is the average over $N_\mathrm{rep}=100$ repetitions of calculations with $N_\mathrm{sample}=1500$ samples plus-minus the corresponding standard deviation, to take into account small fluctuations due to finite sample size. For comparison, the values of $J$ for the initial guess pulse are 0.557, 0.544, 0.482,
    0.394, 0.378 for $\tau_\delta=100,10,1,0.1,0.01$, respectively.}
\end{figure*}

\begin{figure*}
    \vspace{2mm}
    \centering
    \includegraphics[width=\linewidth]{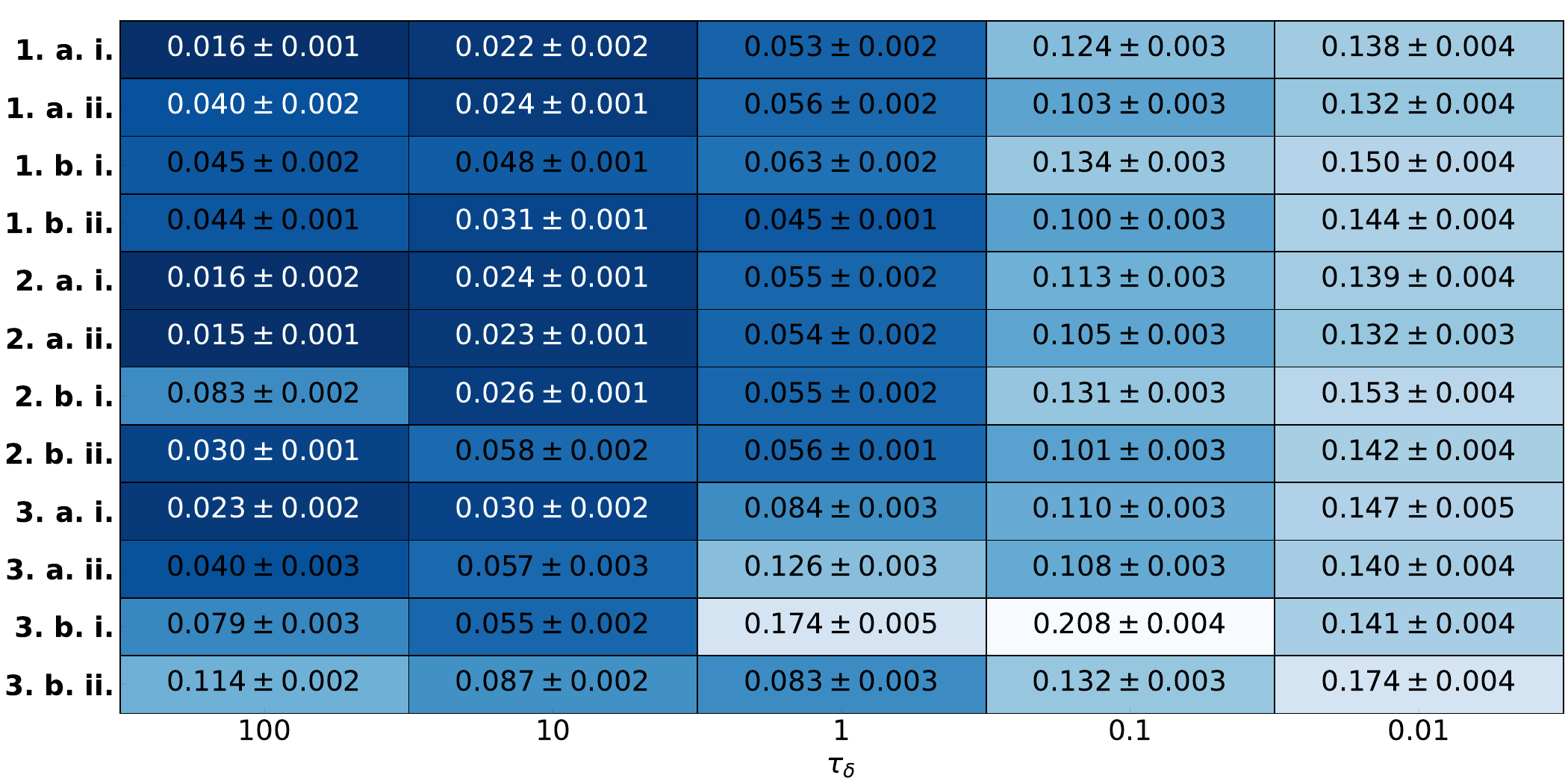}
    \caption{Tile plot of the optimized cost function $J_\mathrm{opt}$ for different $\tau_\delta$ and optimization options. The numbers being printed in white or black are solely for visibility. The optimization option codes are described in Fig.~\ref{fig_main}}
    \label{fig_cf_comparison}
\end{figure*}

\begin{figure*}
    \centering
    \includegraphics[height = 0.8\textheight]{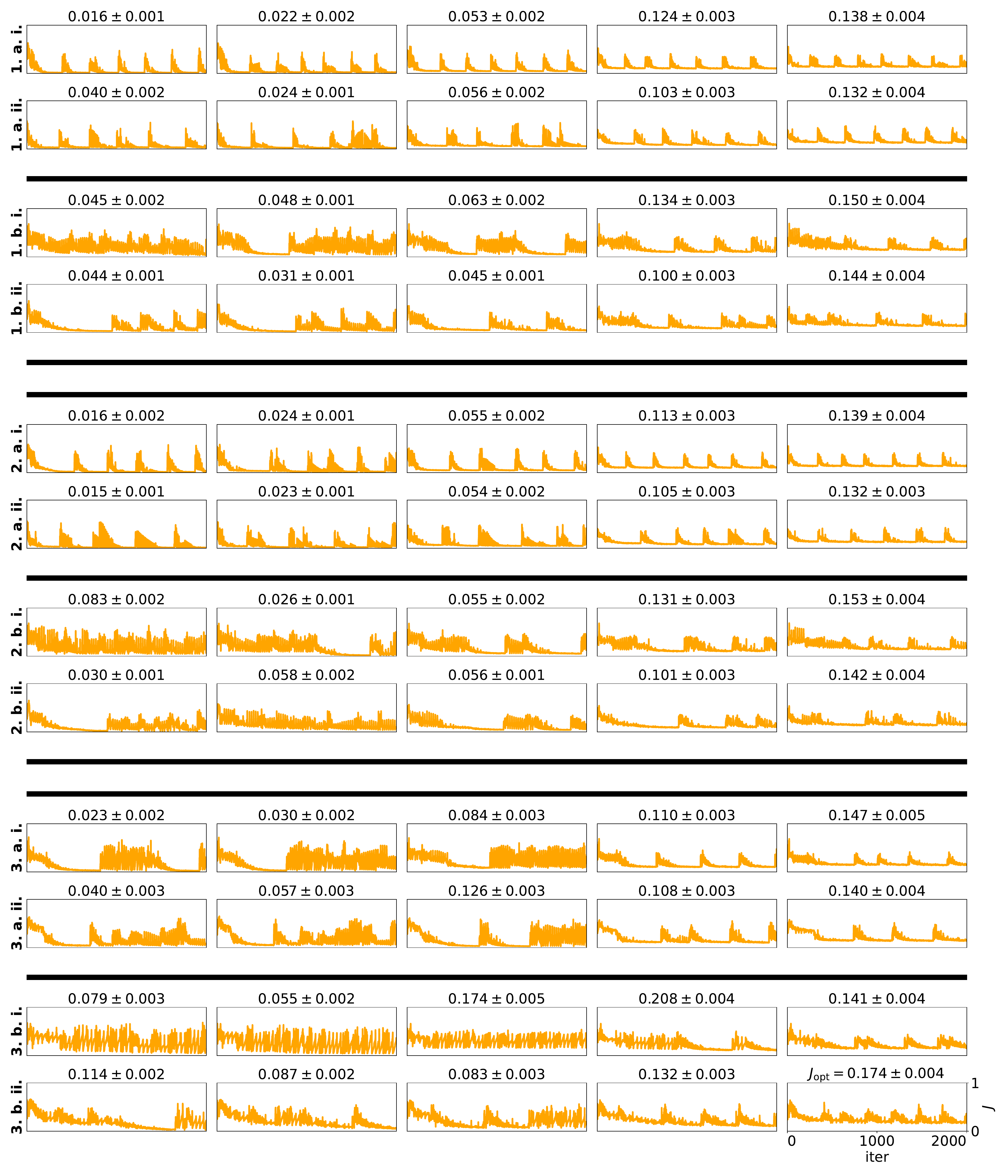}
    \caption{Change in the cost function $J$ over the optimization iterations. A sudden rise in $J$ after converging to some value indicates a superiteration converging and the algorithm moving into the subsequent superiteration. Other features of the plots are the same as Fig.~\ref{fig_main}.}
    \label{fig_cf_evolution}
\end{figure*}

\begin{figure*}\label{fig_main_2}
    \centering
    \includegraphics[height=0.8\textheight]{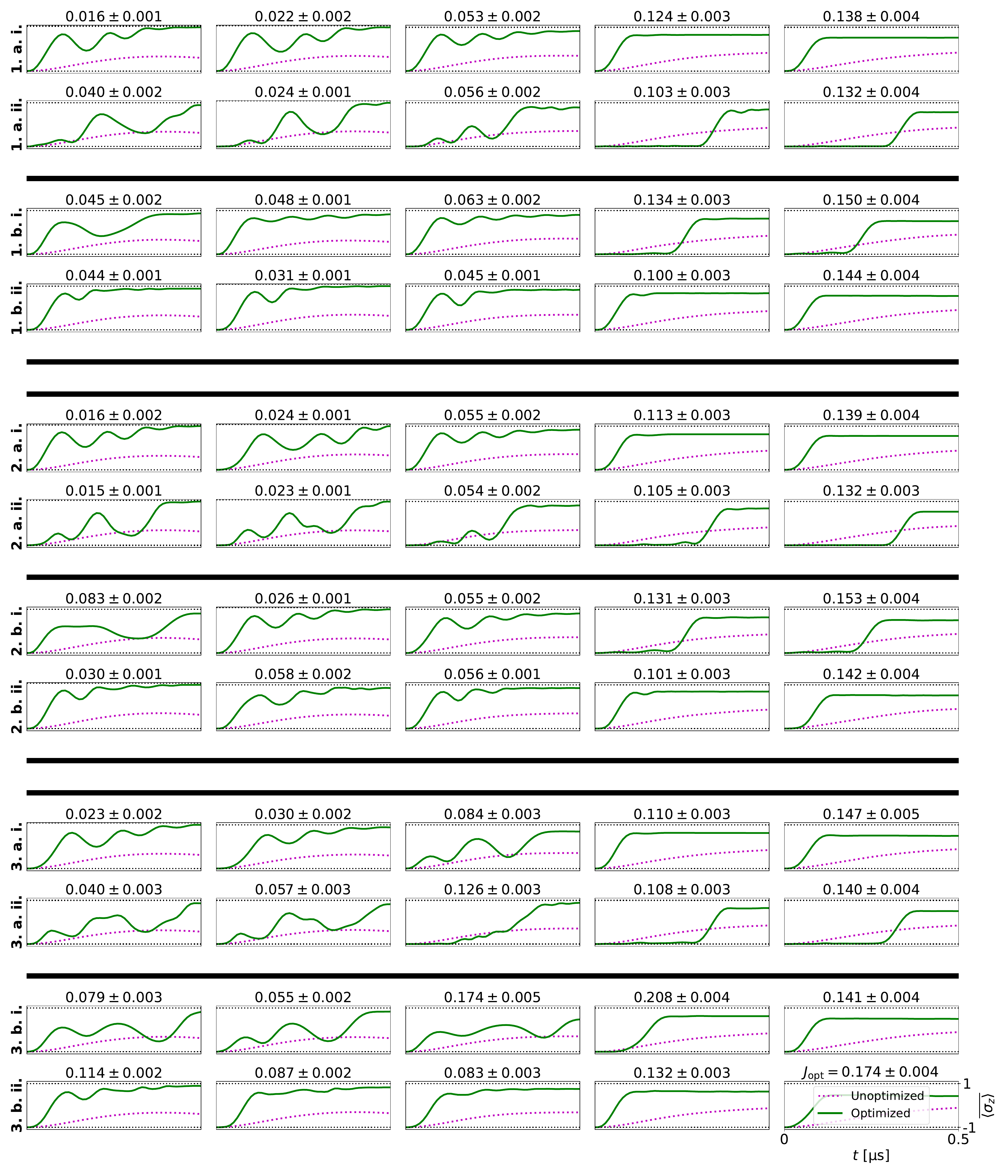}
    \caption{Evolution of $\overline{\expval{\sigma_z}}$ under the optimized pulses presented in Fig.~\ref{fig_main}, compared to the unoptimized version. The green solid lines represent the optimized evolution, while the magenta dotted lines represent the unoptimized evolution. The black dotted lines represent $\pm 1$. Other plot attributes are the same as Fig.~\ref{fig_main}.}
\end{figure*}

We compile all optimization results for the cases discussed in Section \ref{Section_The_Optimal_Control}. Figure~\ref{fig_main} shows the $x$- and $y$- components of the dCRAB-optimized pulses. The comparison of the optimized cost function $J_\mathrm{opt}$ for different cases is compactly presented in Fig.~\ref{fig_cf_comparison}. Meanwhile, Fig.~\ref{fig_cf_evolution} shows the change in the cost function over the optimization iterations. Lastly, Fig.~\ref{fig_main_2} shows the optimized and unoptimized evolution of $\overline{\expval{\sigma_z}}$ representing the population inversion.

\newpage

\end{document}